\documentclass[journal=jacsat,manuscript=letter]{achemso}

\usepackage[version=3]{mhchem}

\newcommand{\TO}{TiO$_2$}

\newcommand{\TiSo}{Ti$_{\textrm{S0}}$}
\newcommand{\PolSo}{Pol$_{\textrm{S0}}$}
\newcommand{\TiSi}{Ti$_{\textrm{S1}}$}
\newcommand{\PolSi}{Pol$_{\textrm{S1}}$}

\newcommand{\eg}{\textit{e.g.},}
\newcommand{\ie}{\textit{i.e.},}
\newcommand{\VO}{V$_{\rm O}$}
\newcommand{\cvo}{c$_{\rm V_{O}}$}
\newcommand{\obr}{O$_{\rm br}$}

\author{Viktor C. Birschitzky}
\email{viktor.birschitzky@univie.ac.at}
\affiliation[UniVie]
{Faculty of Physics and Center for Computational Materials Science, University of Vienna, Vienna, Austria}

\author{Igor Sokolovi\'{c}}
\affiliation[TU]
{Institute of Applied Physics, TU Wien, Vienna, Austria}

\author{Michael Prezzi}
\affiliation[UniVie]
{Faculty of Physics and Center for Computational Materials Science, University of Vienna, Vienna, Austria}

\author{Krisztián Palotás}
\affiliation[UniBud]
{Institute for Solid State Physics and Optics, HUN-REN Wigner Research Center for Physics, Budapest, Hungary}

\author{Martin Setv\'{i}n}
\affiliation[UniPra]
{Department of Surface and Plasma Science, Faculty of Mathematics and Physics, Charles University, Prague, Czech Republic}
\alsoaffiliation[TU]
{Institute of Applied Physics, TU Wien, Vienna, Austria}

\author{Ulrike Diebold}
\affiliation[TU]
{Institute of Applied Physics, TU Wien, Vienna, Austria}

\author{Michele Reticcioli}
\affiliation[UniVie]
{Faculty of Physics and Center for Computational Materials Science, University of Vienna, Vienna, Austria}

\author{Cesare Franchini}
\affiliation[UniVie]
{Faculty of Physics and Center for Computational Materials Science, University of Vienna, Vienna, Austria}
\alsoaffiliation[UniBo]
{Dipartimento di Fisica e Astronomia, Università di Bologna, Bologna, Italia}

\title[Polaron ML]
{Machine Learning Based Prediction of Polaron-Vacancy Patterns on the TiO$_2$(110) Surface}

\begin{document}

\begin{abstract}
    The multifaceted physics of oxides is shaped by their composition and the presence of defects, which are often accompanied by the formation of polarons. 
    The simultaneous presence of polarons and defects, and their complex interactions, pose challenges for first-principles simulations and experimental techniques.
    In this study, we leverage machine learning and a first-principles database to analyze the distribution of surface oxygen vacancies (\VO) and induced small polarons on rutile TiO$_2$(110), effectively disentangling the interactions between polarons and defects. 
    By combining neural-network supervised learning and simulated annealing, we elucidate the inhomogeneous \VO\ distribution observed in scanning probe microscopy (SPM).
    Our innovative approach allows us to understand and predict defective surface patterns at previously inaccessible length scales, identifying the specific role of individual types of defects.
    Specifically, surface-polaron-stabilizing \VO-configurations are identified, which could have consequences for surface reactivity.

\end{abstract}

\section{Introduction}

The rich and tunable physics of oxides depend on their precise chemical composition, and the presence of impurities, including atomic vacancies, interstitial atoms, and dopants in the material~\cite{rousseau_theoretical_2020, Franceschi2022, Jupille2015book, strand_structure_nodate}.
Defects at the atomic level frequently lead to the formation of polarons, which are localized charge carriers arising from the synergy between unbound charges and lattice phonons~\cite{Franchini2021, Emin2013, Alexandrov2010}.
In the specific case of so-called small polarons, the polaronic charge is localized almost entirely on one atomic site, surrounded by sizable distortion of the local lattice structure~\cite{Stoneham2007}.
In conjunction with their inducing defects, these small polarons play a dominant role in a wide range of processes relevant to technological applications~\cite{pastor_electronic_2022} and fundamental phenomena such as charge carrier mobility~\cite{kick_mobile_2020, Chen2023, Smart2017}, electron-hole recombination\cite{cheng_co_2022, cheng_photoinduced_2022} and adsorption\cite{Sokolovic2020, tanner_polaron-adsorbate_2021, yim_visualization_2018}.

Most importantly the role of polarons is known to be highly relevant in the context of (photo)catalysis~\cite{run_photo_2023,ren_recent_2023,dohnalek_thermally-driven_2010, tanner_tio2_2022} and, single-atom catalysis~\cite{sombut_role_2022, geiger_coupling_2022, geiger_data-driven_2022}.
The localized charge carriers act as active centers, which enhance (photo)catalytic activity by providing sites that can readily adsorb and interact with reactant molecules~\cite{cao_scenarios_2017, reticcioli_interplay_2019}.
Although polaron formation may in principle occur on any site of the lattice, the defects can act as attractive or repulsive centers, favoring specific polaronic configurations over others~\cite{birschitzky_machine_2022}.
In turn, the dynamics and distribution of the atomic defects are known to be altered by the polarons~\cite{Zhang2019}.
Therefore, control over the spatial distribution of polaronic active centers becomes pivotal in optimizing (photo)catalytic performance.

While theoretical studies based on density functional theory (DFT) have elucidated excess charge localization in relation to the inducing defect in many materials~\cite{ellinger_small_2023, osterbacka_charge_2022, Sun2017}, the specific role of subsurface and surface polarons, particularly in the presence of defects, on the archetypal redox active oxide surface \TO(110) is still debated.
Here, a problem arises from the complexity of the configuration space of point impurities, where DFT calculations strive to account for the computational cost of the problem.
As a consequence, either no exploration attempt is performed (\ie\ most studies rely on the configuration randomly obtained in the DFT calculation)~\cite{Reticcioli2019b, Pham2020}, or effective but costly approaches are adopted such as molecular dynamics~\cite{Reticcioli2017d, Kowalski2010}, Monte-Carlo-driven DFT simulations~\cite{Han2018} or systematic explorations limited to a handful of localization sites~\cite{reticcioli_formation_2018}.
Thus, finding a method that effectively navigates the diverse defect-polaron configuration landscape has become a research imperative.

In this study, we focus on rutile \TO(110) and show how the spatial distribution of \VO\ measured by SPM can be successfully predicted and interpreted by first-principles calculations if the coupling between \VO\ and polarons is taken into account.
To address this problem, we developed a strategy based on defect distribution descriptors and neural networks to predict the stability of specific polaron-vacancy patterns.
Through an iterative optimization active learning cycle, we systematically extended the DFT reference dataset and converged the machine learning (ML) model, to efficiently explore the defect-polaron configuration space.
The model can capture the complexity of the \VO-polaron interactions with DFT accuracy and proposes new configurations showing remarkable energy stability.
By feeding Markov-chain Monte-Carlo (MC) algorithms with the ML configuration energies, we simulate the annealing process leading to the formation of vacancies and polarons in the experimental samples.
As a final result, we obtain large-area (\textgreater10$\times$10$\,\text{nm}^2$) surface morphologies resembling the SPM measurements.
This analysis revealed new physical properties of the polarons on \TO(110), where the formation of inhomogeneously distributed \VO\ is linked to an increased formation of surface polarons and, therefore, to the density of active sites.

\section{Results and Discussion}

\subsection{Defect Distribution via DFT, Experiment, and Machine Learning}

    \begin{figure}[ht]
        \includegraphics[width=\textwidth]{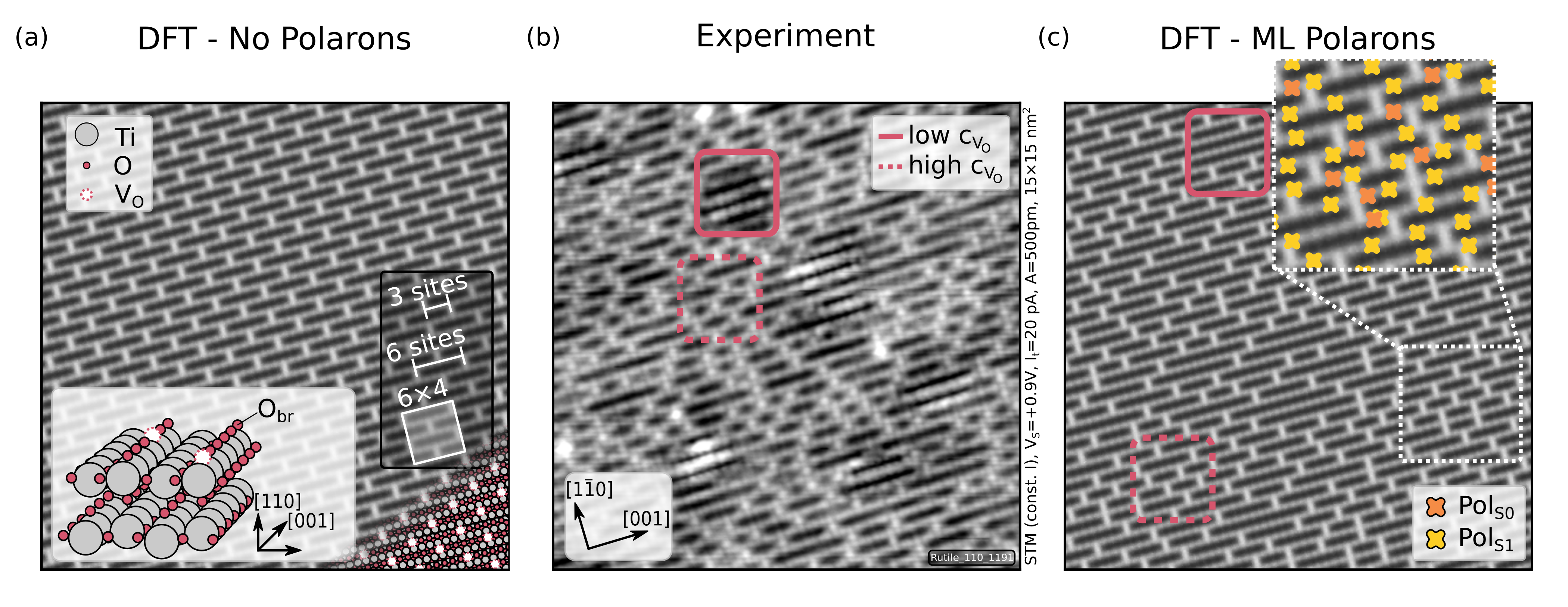}
        \caption{Oxygen vacancy distribution on rutile TiO$_2$(110) obtained by various methods. 
        (a) Schematic representation of the most favorable \VO\ distribution in non-polaronic DFT calculations as obtained from a 6$\times$4 ($\sim$1.8$\times$2.6$\,\text{nm}^2$) supercell. 
        The schematic depiction is generated by showing the \obr\ bridging atoms as black regions and Ti$_\mathrm{5c}$ rows and \VO\ as white.
        The inset displays the structural model of rutile \TO(110). 
        The distance maximizing \VO\ distribution (6 sites in row, 3 sites in adjacent row) and the 6$\times$4 supercell are indicated.
        (b) Unoccupied-states, constant-current STM image of a clean, reduced rutile TiO$_2$(110) surface (imaging parameters in the Figure) depicting Ti$_\mathrm{5c}$ rows and V$_\mathrm{O}$s as bright, while O$_\mathrm{br}$ rows are depicted as dark. 
        More details on the contrast formation are given in the Methods.
        Locally low and high \VO-concentrations (\cvo) areas are marked with solid and dashed red boxes, respectively.
        The crystalographic directions are consistent in all panels.
        (c) ML-predicted schematic representation of surface oxygen vacancy distribution, where the interaction of surface and subsurface polarons (\PolSo\ and \PolSi, respectively) and \VO s are modeled in a 54$\times$24 ($\sim$ 16$\times$16$\,\text{nm}^2$) supercell.
        Orange and yellow markers show the position of surface and subsurface polarons in the ML prediction.}
        \label{fig:STM}
    \end{figure}

Fig.~\ref{fig:STM} shows the surface structure of reduced rutile \TO(110) as imaged by constant current STM measurements (see panel b and Methods Section), together with the models predicted from DFT without taking polarons into account or by explicitly modeling their impact via machine learning (see panels a and c respectively).
The unreconstructed 1$\times$1 rutile surface consists of alternating rows of under-coordinated (two-fold) oxygen atoms (the bridging oxygen atoms, \obr) and five-fold coordinated titanium atoms (Ti$_\mathrm{5c}$) running along the [001] direction~\cite{Diebold2003, setvin_direct_2014}.
Oxygen vacancies form easily on the \obr\ sites upon sputtering and annealing, up to a critical concentration of \cvo$\simeq17$\%~\cite{Reticcioli2017d}.
At stronger reducing conditions, the surface undergoes a structural reconstruction\cite{Onishi1994, Li1999a, Li2000, McCarty2003, Wang2014a, Mochizuki2016}.
Every \VO\ releases two excess electrons that form polaronic states, localizing preferably on subsurface Ti sites~\cite{birschitzky_machine_2022, Deskins2009, Kowalski2010}.
Thus, the \VO\ can be considered as a positively charged ($2+$) center.
By simple electrostatic considerations (and by, simultaneously, neglecting the role of polarons), one would expect a purely repulsive interaction among the vacancies.
In this picture, the configuration maximizing the \VO-\VO\ distances represents the most favorable vacancy distribution.
For the critical concentration of \cvo$=17$\%, this corresponds to a homogeneous configuration with a \VO-\VO\ distance of 6 lattice sites along the [001] row, and 3 lattice sites considering two oxygen vacancies on adjacent rows (see Fig.~\ref{fig:STM}a).

DFT calculations confirm the homogeneous \VO\ distribution in Fig.~\ref{fig:STM}a as the ground state configuration, as far as the formation of the polarons is suppressed (\ie\ the excess electrons are forced into spatially delocalized states at the bottom of the conduction band, rather than localized polaronic states). 
While this unphysical metallic solution (rutile TiO$_2$ is an n-type semiconductor) is less stable than the polaronic solution, it simplifies the search for the optimal defect distribution via a two-step process. 
Initially identifying the optimal defect pattern through DFT calculations, where polaron formation is suppressed, and subsequently introducing polarons into random positions or finding the most favorable polaron configuration within the given defect distribution~\cite{birschitzky_machine_2022}. 
While this approach reduces the combinatorial divergence of defect-polaron configurations, it relies on the assumption that the distribution of atomic defects is not affected by the polarons, which is not valid for most materials~\cite{Zhang2019}.

The experimental measurements do not support such homogeneous \VO\ pattern.
Fig.~\ref{fig:STM}b shows a typical image as obtained from low-temperature STM measurements on a \TO(110) surface after sputtering and annealing treatment to form a high content of oxygen vacancies (\cvo$\simeq14$\%, close to the critical value of $17$\%).
At this temperature, the oxygen vacancies (imaged as bright spots along the dark [001] \obr\ rows) are immobile and appear in irregular patterns, quite far from any homogeneous distribution.
The discrepancy with the simple models discussed above is a strong indication of the role that polarons can have in determining the optimal \VO\ surface structure.
Simply adding polarons on a rigid \VO\ pattern (effectively decoupling \VO\ and polaron) as usually done in standard DFT simulation, would not improve the situation. 

Fig.~\ref{fig:STM}c reports the surface structure as predicted by our machine learning model, which allows simultaneously varying both \VO\ and polaron positions to find the configuration that minimizes the total energy of the system.
The resulting \VO\ distribution is in good qualitative agreement with the inhomogeneous distribution found throughout experiments.
Our methodology, described in detail in the following, is capable of capturing the effects of the polarons on the oxygen vacancy distribution, going beyond the simple picture relying on purely \VO-\VO\ interactions.
Moreover, it allows us to consider large surface areas of about 250$\,\text{nm}^2$ (\textgreater15$\times$15$\,\text{nm}^2$), corresponding to 54$\times$24 supercells, extending considerably the limits of standard DFT simulations.

\subsection{Machine Learning Polaron and Defect Distributions}

The methodology proposed here is structured in three parts:
First, we train a feed-forward neural network~\cite{lecun_deep_2015} to predict the DFT energy of the system depending on the configurations of the impurities.
Due to the computational limitation of the DFT calculations, we adopt relatively small unit cells in this step.
Specifically, we used two supercells with different lateral extensions (6$\times$4 and 12$\times $2, see Methods Section) to include long-range interactions along different crystallographic directions.
Then, we use the trained model to search for low-energy configurations that were not included in the original set of data, adopting an active learning scheme~\cite{behler_four_2021}.
Finally, we use the actively trained model to obtain large-area predictions.
In the following, we describe in detail the architecture of the machine-learning model and compare the ML predictions with experimental data on reduced \TO(110).

The training of the machine learning model requires a reference database built up by several, distinct polaron and atomic-defect configurations.
By following the process described in detail in the Methods Section, we calculated the free energy for different configurations at the DFT+\textit{U} level using VASP~\cite{kresse_efficiency_1996, kresse_ultrasoft_1999, blochl_projector_1994}, with a \textit{U}$=3.9$~eV on the $d$ orbitals of Ti atoms~\cite{Dudarev1998, birschitzky_machine_2022}.
Polarons were localized at chosen surface \TiSo\ (\PolSo) and subsurface \TiSi\ (\PolSi) sites via occupation matrix control~\cite{allen_occupation_2014}.
We modeled 2367 symmetrically-inequivalent polaron-\VO\ configurations in a 6$\times$4 supercell (\ie\ six and four times the [001] and [1$\bar{1}$0] lattice vectors, respectively), and 2155 configurations in a 12$\times$2 supercell.
To optimize the model, we randomly split the calculated configurations and energies into training and validation data sets, including 80\% and 20\% configurations, respectively.

    \begin{figure}[ht]
        \centering
        \includegraphics[width=\textwidth]{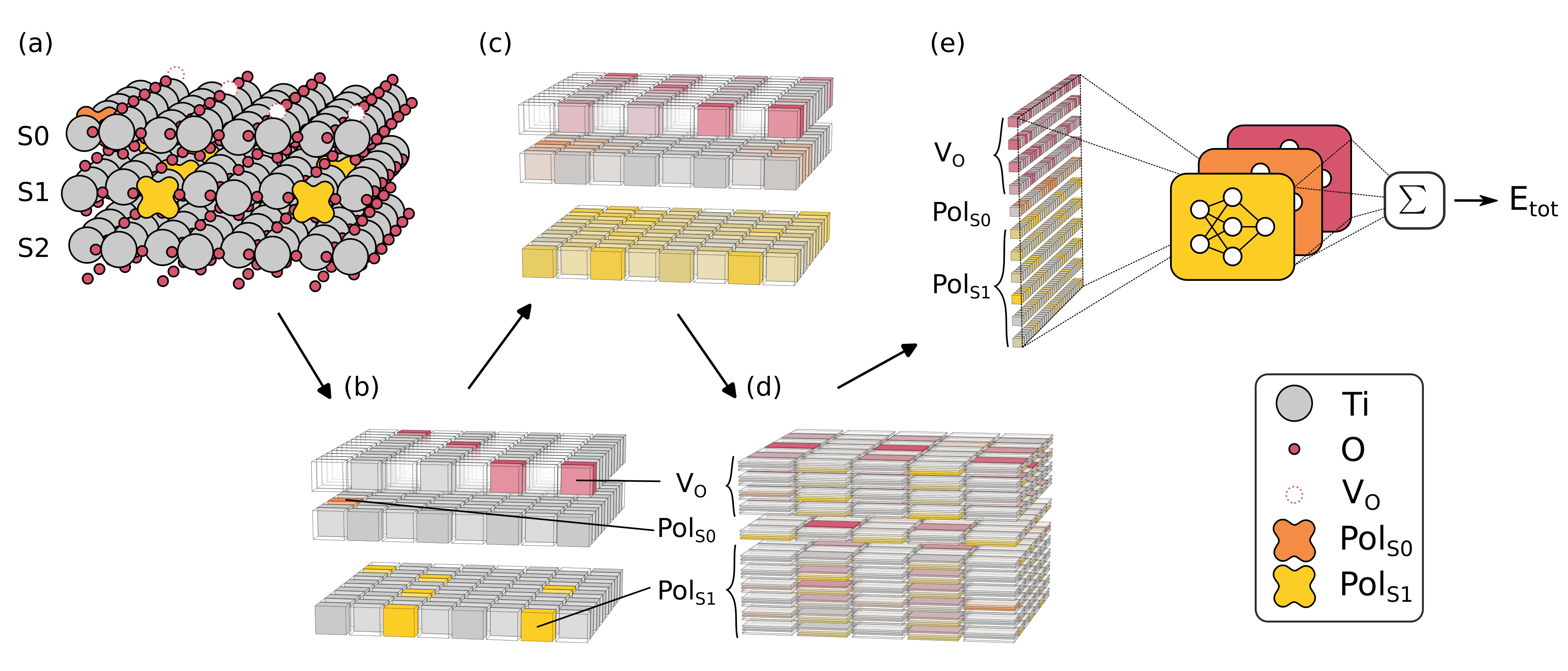}
        \caption{Machine learning model architecture. (a) A defect structure consisting of oxygen vacancies and polarons in a supercell. (b) The supercell is converted into a discretized grid, where each cell encodes whether it contains a defect/polaron. (c) Smearing of the one-hot encoding. (d) The supercell is partitioned into the local environment of each defect. (e) The local environment descriptors are fed through a feed-forward neural network to predict the energy contribution of each defect. The sum of the individual defect contributions gives the total energy of the system.}
        \label{fig:ML}
    \end{figure}

Fig.~\ref{fig:ML} sketches the main features of our ML architecture predicting the stability of different defect-polaron configurations
A generic polaron-\VO\ distribution on the rutile TiO$_2$ surface is depicted in Fig.~\ref{fig:ML}a. 
The descriptor representing the configuration is constructed by, first, discretizing the space into a rough grid (see Fig.~\ref{fig:ML}b), and encoding the spatial distribution of polarons and atomic defects.
The discretized space simplifies the training of the ML model, as compared to using Euclidean distances~\cite{birschitzky_machine_2022}.
To improve the description of spatially close polarons/defects, we employ a one-hot encoding (\ie\ value of 1 for grid cells containing a defect, 0 otherwise), smeared via multiple applications of a discrete Laplacian kernel (Fig.~\ref{fig:ML}c).
Then, to predict the energy of the whole system given a specific configuration, we split the total energy into contributions arising from a single defect/polaron impurity (Fig.~\ref{fig:ML}d):
    \begin{equation}
        E_\textrm{tot} = \sum_{i}^{N_\textrm{S1}}E_i +\sum_{j}^{N_\textrm{S0}}E_j + \sum_{k}^{N_\textrm{VO}}E_k ~~\textrm{.}
    \end{equation}
Here, $E_\textrm{tot}$ is the total energy of a given configuration, and $E_i$, $E_j$ and $E_k$ are the virtual contributions of a single \PolSi, \PolSo, and \VO\ respectively.
We use a feed-forward neural network to estimate the virtual contribution of a single defect/polaron (Fig.~\ref{fig:ML}e).
Finally, we sum over the virtual contributions to obtain the total configuration energy~\cite{birschitzky_machine_2022}.
The total energy can be computed by DFT calculations~\cite{reticcioli_formation_2018}, while the virtual contributions are not directly accessible in the DFT data.
Thus, we can train our model using the discretized defect-polaron positions as a descriptor, and the DFT energy as the target quantity.
By training the ML model on DFT data obtained for the 6$\times$4 unit cell (see SM Fig.~S1), we achieved a mean absolute error of $1.8$ and $2.2$~meV/\VO\ for the training and validation sets, respectively.
By adding training data from the 12$\times$2 unit cell (see SM Fig.~S2), the mean absolute error increased slightly ($2.9$ and $3.5$~meV/\VO\ in training and validation, respectively).
However, by using both sets of data in the training, the ML model can account for longer interactions in both the [$001$] and [$1\bar{1}0$] directions.
For a detailed description of the training process see Methods Section.

Aiming for a comparison with the experimental measurements, we focus here on the low-energy configurations, which are more likely to get stabilized in real samples.
To identify such stable configurations, we performed simulations that model the annealing process.
In the preparation of the experimental samples, both polarons and oxygen vacancies diffuse on the sample during annealing.
At lower temperatures, \VO\ on rutile are immobile, while polarons always show a certain degree of mobility, hopping/tunneling a few lattice sites around the equilibrium position~\cite{setvin_direct_2014, Reticcioli2017d}.
The simulated annealing can be implemented as a global optimization scheme~\cite{kirkpatrick_optimization_1983}.
Candidate configurations are obtained by perturbing the current configuration, randomly displacing one defect/polaron impurity to any nearest neighbor site.
The new configuration is either accepted or declined by virtue of the Metropolis-Hastings algorithm~\cite{metropolis_equation_2004} with the acceptance criterion based on the configuration energy -- similar in spirit to large-scale defect distribution studies based on reverse MC~\cite{ji_oxygen_2022} (although in our approach the defect distribution is not fitted to minimize the deviation from experiment, but it relies entirely on DFT/ML data).
Calculating the energy of the candidate configurations within the DFT framework would make this approach unfeasible, due to the computational cost of DFT calculations and the high number of energy evaluations required for a single optimization.
Conversely, the ML model allows us to inspect the stability of an extremely high number of defect-polaron configurations (minimization of the energy requires on the order of $10^3-10^6$ energy evaluations depending on the size and initialization of the employed configuration) and enables the exploration of candidate structures.

We iterated annealing simulations following an active learning procedure.
The initial DFT data set built by random configurations was progressively augmented by including the results from the annealing optimization (see SM Fig.~S3).
Consequently, we obtained a final ML model refined to account for a broader range of configurations.
The refined model is finally used to obtain large-area predictions (7.1$\times$10.5$\,\text{nm}^2$ to collect statistics and 16$\times$16$\,\text{nm}^2$ for visualizations) on the defect-polaron distributions, using again the simulated annealing approach.
The qualitative agreement with the experimental data is shown in Fig.~\ref{fig:STM}c for the 54$\times$24 rutile TiO$_2$(110) supercell. 
In the following, we quantitatively analyze our results.

\subsection{Formation of \VO-Polaron Patterns and Their Mutual Interaction}

    \begin{figure}
        \centering
        \includegraphics{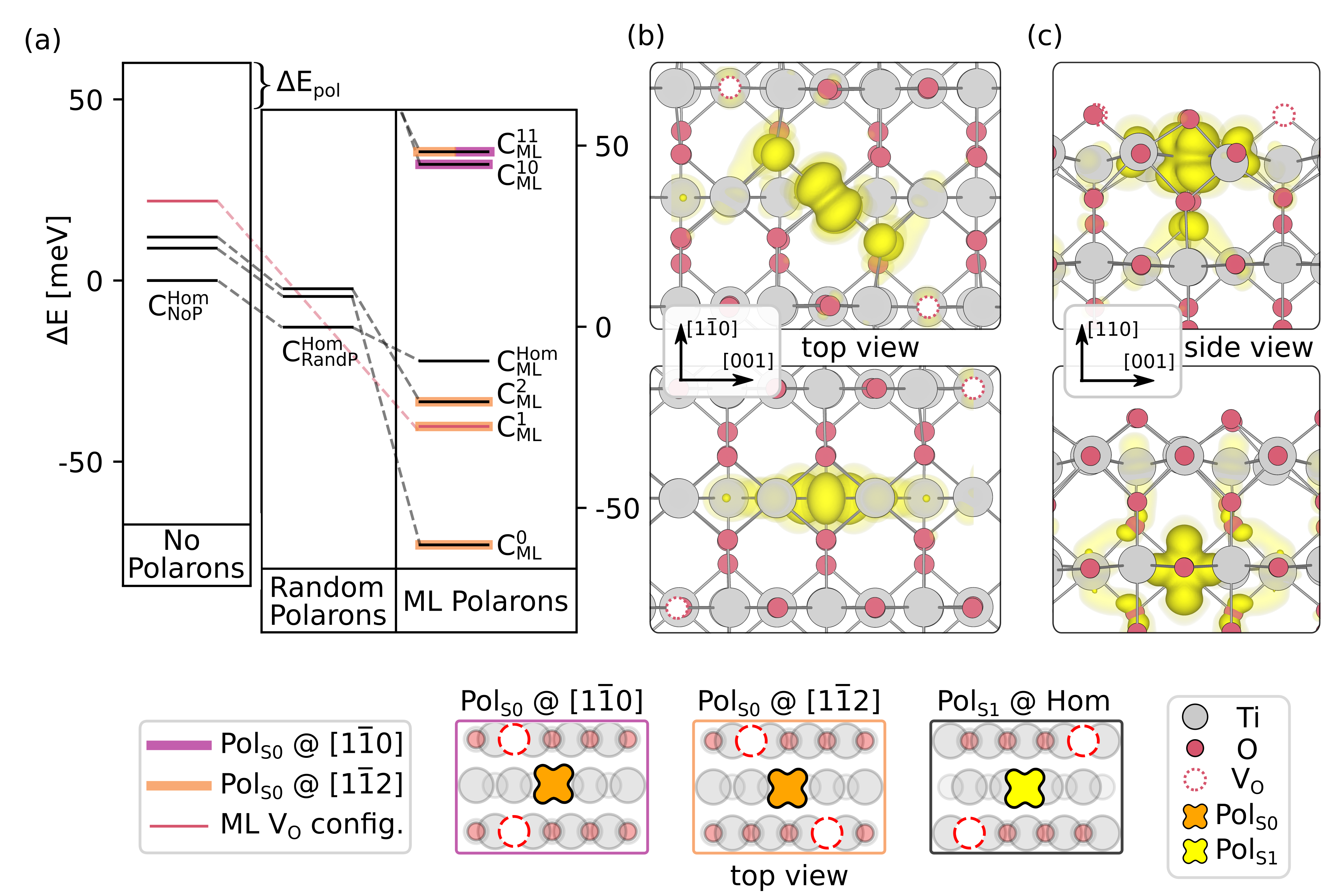}
        \caption{Analysis of \VO-polaron configurations in a 6$\times$4 \TO(110) cell. 
        (a) Comparison of selected low-energy vacancy-polaron configurations as obtained by different treatments of the polaron-\VO\ interaction. 
        For a comparison of all configurations and their labeling, refer to Supplementary Fig. S5.
        The change in energy for all low-energy configurations is displayed in Supplementary Fig. S6.
        ``No Polarons'' refers to DFT calculations suppressing polaron formation.
        ``Random Polarons'' refers to the reference DFT data set, built by including polarons in random positions or guided by physical intuition.
        ``ML Polarons'' indicates the DFT energies of configurations identified in the ML search.
        Total energies $\Delta E$ are shown using the homogeneous \VO\ distributions (from ``No Polarons'' and ``Random Polarons'') as references (note the large energy gain $\Delta E_{\mathrm{pol}}$ of -3.23\,eV between non-polaronic and polaronic solutions with homogeneous \VO\ patterns).
        Dashed lines connect identical \VO~configurations. 
        New \VO~configurations found in the ML search are displayed in red.
        The occurrence of \TiSo\ polarons is highlighted in orange and purple for [1$\bar{1}$2]- and [1$\bar{1}$0]-aligned oxygen vacancies, respectively.
        The most important \VO-polaron complexes are shown schematically in top view at the bottom of the Figure.
        Only the most stable polaronic configuration per \VO\ arrangement is shown.
        (b,c) Top and side views of the polaronic isocharge surfaces of the [1$\bar{1}$2]-aligned \VO-\PolSo\ complex (top), and of the \PolSi\ in the homogeneous \VO-distribution (bottom).
        }
        \label{fig:surface_pol}
    \end{figure}

The analysis of the low-energy configurations (see SM Fig.~S4 for the energy distribution of all possible \VO\ configurations in the 6$\times$4 cell) is summarized in Fig.~\ref{fig:surface_pol}.
Fig.~\ref{fig:surface_pol}a shows the improvement of energies of the \TO(110) 6$\times$4 cell as obtained by treating \VO-polaron coupling at three different levels: (i) Suppressing polarons (``No Polarons''); (ii) Distributing polarons in random or positions guided by physical intuition (``Random Polarons'' ); (iii) Full inclusion of polaron-\VO\ interaction via our proposed ML protocol (``ML Polarons'').
By suppressing polaron formation, the ground state configuration is given by the vacancies being homogeneously distributed on the surface (C$_{\rm NoP}^{\rm Hom}$ configuration, see ``No Polarons'' column in Fig.~\ref{fig:surface_pol}a).
The ``Random Polarons'' column of Fig.~\ref{fig:surface_pol}a shows instead the energy of the system obtained by including polarons in random positions and enriched by adding specific, low-energy polaronic configurations that were suggested in previous studies~\cite{reticcioli_formation_2018, Reticcioli2017d, birschitzky_machine_2022}.
Here, the ground state configuration is given by a homogeneous distribution of \PolSi\ in the homogeneous \VO\ pattern (labeled as C$_{\rm RandP}^{\rm Hom}$ in Fig.~\ref{fig:surface_pol}a).

Treating polaron-\VO\ coupling at the ML level (third approach) results in novel \VO\ distributions with lower energy, indicating a new ground state for the system, where the homogeneous configuration is no longer the most stable one, as shown in the ``ML Polarons'' column of Fig.~\ref{fig:surface_pol}a.
First, we note that the ML model identified a different order of \PolSi\ showing a better stability in the homogeneous \VO\ background (labeled as C$_{\rm ML}^{\rm Hom}$, see also SM Fig.~S5).
Moreover, new polaron configurations explored by the extensive ML search improve the stability of many other \VO\ patterns (see the energy levels in black in the ``ML Polarons'' column of Fig.~\ref{fig:surface_pol}a, lower than in the ``Random Polarons'' column).
Importantly, two of these previously-unexplored polaron configurations (labeled as C$_{\rm ML}^{\rm 0}$ and C$_{\rm ML}^{\rm 2}$) resulted in energy values even lower than the homogeneous distribution, revealing a new ground state for the system.
Moreover, novel \VO-patterns were proposed by the ML search as low-energy configurations.
One in particular (red line in Fig.~\ref{fig:surface_pol}a) is ranked as the second most stable configuration (C$_{\rm ML}^{\rm 1}$).
The polarons play a key role in stabilizing this \VO-pattern and as further proof, we calculated the energy of this new \VO-pattern, artificially suppressing the polaron formation, and obtained a much worse stability (red line in the ``No Polarons'' column).

Interestingly, in all the new low-energy configurations 
obtained from the ML-driven search (except for C$_{\rm ML}^{\rm Hom}$), we note the presence of at least one polaron on a surface \TiSo\ site (configurations containing \PolSo\ are orange highlighted in Fig.~\ref{fig:surface_pol}a).
Fig.~\ref{fig:surface_pol}b and c compare the spatial distribution of the surface \PolSo\ and subsurface \PolSi.
The formation of the surface polaron is particularly stable when occurring in the central \TiSo\ site between two oxygen vacancies aligned on the [1$\bar{1}$2] direction (see top view in Fig.~\ref{fig:surface_pol}b).
This [1$\bar{1}$2]-aligned \VO-polaron complex represents indeed the ground state configuration obtained by our ML search (\eg\ it is present in C$_{\rm ML}^{\rm 0,1,2}$).
Another remarkably stable complex is given by two vacancies aligned along the [$1\bar{1}0$] direction and one \PolSo\ in their vicinity (see SM Fig.~S5).
For instance, this complex appears in the configurations highlighted in purple in Fig.~\ref{fig:surface_pol}a (C$_{\rm ML}^{\rm 10,11}$).
The [1$\bar{1}$2] and [$1\bar{1}0$] alignments found in the ML search agree well with the experimental SPM measurements (compare Fig.\ref{fig:STM}b and c) showing a high coverage of such high density \VO-regions. 
In contrast, DFT predictions, which neglect polaron-\VO\ interaction or randomly distribute polarons, favor homogeneous configurations.

\subsection{Comparison of a Large Scale Model and the Experimental Surface}

\begin{figure}
        \centering
        \includegraphics{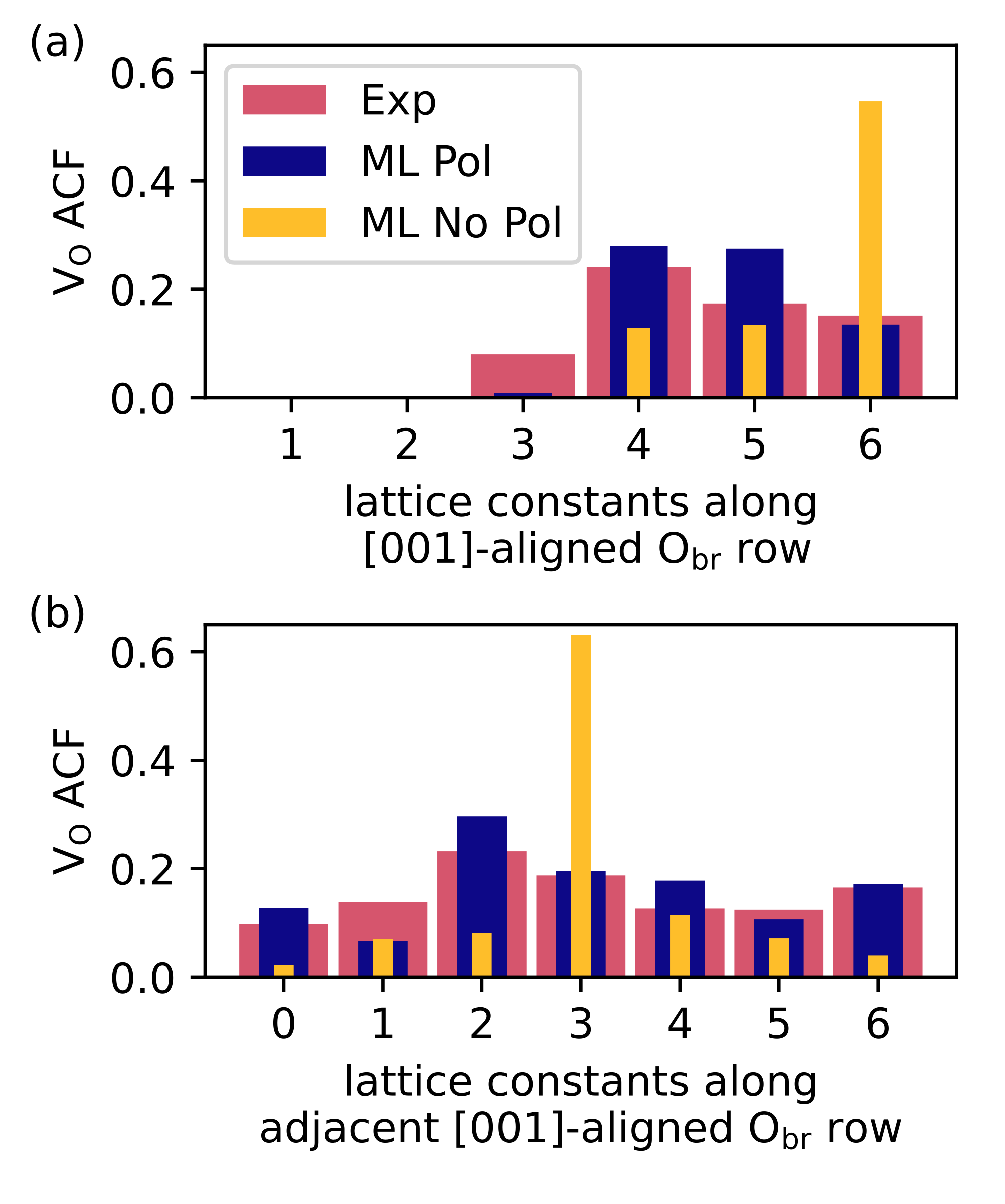}
        \caption{Autocorrelation functions of the \VO\ positions as extracted from Fig.~\ref{fig:STM}b (Exp) and ML-based ACFs with (ML Pol) and without polarons (ML No Pol).
        (a,b) Comparison of experimental and simulated \VO\ autocorrelation functions along a single (a) and adjacent (b) [001]-aligned \obr\ row. 
        Experimental autocorrelation functions are averaged to remove remaining anisotropies.
        Simulated ones are averaged over all symmetrically equivalent most stable configurations from 60 differently seeded simulated annealing runs performed in 24$\times$16 supercells.
        ML Pol and ML No Pol are started from identically seeded \VO\ patterns, with 2$N_{\mathrm{V_O}}$ polarons and no polarons, respectively.
        Autocorrelation functions are rescaled to account for \cvo\ differences in experiment (14.2\%) and simulation (16.7\%).
        }
        \label{fig:dist}
    \end{figure}

Fig.~\ref{fig:dist} shows our results as obtained by ML-driven annealing simulations on large-area 24$\times$16 cells (corresponding to 7$\times$10$\,\text{nm}^2$), which enables a direct comparison with the experiment.
Visual inspection (Fig. \ref{fig:STM}b vs. c) already indicates that our ML treatment provides \VO\ distribution that closely resembles the experimental one. 
We quantify this agreement by calculating autocorrelation functions (ACF)\cite{Setvin2015b} for simulated annealings under different computational conditions and compare it to the experimental ACF of the \VO\ distributions extracted from Fig. \ref{fig:STM}b (for details see Supplementary Fig.~S7).
The simulated annealing procedure starts from random \VO-polaron configurations, where we obtain several large-area models (such as the one in Fig.~\ref{fig:STM}c), all showing very similar characteristics.
To complete our comparison, we also use the ML model to anneal a system where polaron formation is suppressed.
This scheme, similar to the non-polaronic DFT approach of Fig.~\ref{fig:STM}a, assumes a homogeneous \VO\ pattern but takes into account annealing-induced disorder effects.

The ACFs are shown in Fig.~\ref{fig:dist}, where projections of \VO\ defect populations along the same and adjacent [001] rows are shown in the histograms in panels a and b, respectively.
For oxygen vacancies lying on the same row, both the ML model and the experiments show that short \VO-\VO\ distances of 1 and 2 lattice sites are unlikely.
The highest probability lies at a distance of 4 or 5 lattice sites for both the experimental and ML annealing including polaron-\VO\ interactions (see Fig.~\ref{fig:dist}a and b red and blue data, respectively).
By considering only the \VO-\VO\ repulsion as driving force (\ie\ excluding polaron formation in the ML annealing procedure; see ML No Pol in Fig.~\ref{fig:dist}a and b) and applying an identical annealing protocol as in the polaron-\VO\ interaction case, we find the probability maximum lying at a 6-site distance for in-row and 3-site distance in the adjacent row.
This is further evidence for polarons' role in stabilizing the \VO\ arrangement.

As a result, the rutile TiO$_2$ surface shows some areas with a locally low density of oxygen vacancies (down to 0\%), alternated with highly dense areas (up to 20\%, which is compatible with the 4-site-distance distribution).
Our data suggest that the great stability of the [1$\bar{1}$2]- and [$1\bar{1}0$]-aligned \VO-polaron complexes contribute to this alternation of locally less and more reduced areas at a given \cvo.\ 
To further corroborate this result, we performed additional DFT calculations modeling this strong inhomogeneity (see SM Fig.~S5 with configurations C$_{\rm ML}^{\rm 26-29}$).
We also note that this analysis reconciles the DFT predictions on the critical concentration at which the (1$\times$2) surface reconstruction occurs for the surface phase transition, which was calculated as $\sim 20\%$, in apparent disagreement with the experiments reporting an average concentration of $17$\%~\cite{Reticcioli2017d}.

\section{Conclusions}
In summary, we directly elucidated the impact of polarons on the structure of oxide surfaces, using an example of the prototypical rutile TiO$_2$(110) surface.
Specifically, we designed a computational machinery to predict the distribution of polarons and oxygen vacancies on rutile \TO(110), by performing machine-learning-guided DFT calculations.
MC-driven annealing simulations based on the ML data enabled the exploration of defect distributions on scales much larger than standard DFT allows.
An analysis of the experimental SPM images yielded a direct validation of the theoretical predictions.
While conventional approximations used in traditional DFT calculations result in homogeneous solutions, we were able to retrieve the inhomogeneity of the \VO\ distribution as detected by the experiments.
Our analysis clarifies the peculiar inhomogeneous distribution of \VO\ on rutile \TO(110).
Most importantly, the system shows a tendency towards the formation of high-density \VO\ patterns, alternated with low-density \VO\ regions.
While larger defect-free areas are typically attributed to subsurface Ar impurities~\cite{doi:10.1021/nl5024194}, the here observed fluctuation of the local \cvo\ can partially be attributed to the interaction of polarons and \VO s.

These results suggest that surface reactivity could be optimized by tuning the annealing procedure to facilitate the formation of energetically more favorable, high-density \VO\ patterns, which promote surface localized charges and their interaction with adsorbates~\cite{reticcioli_interplay_2019}.
To elucidate the role of the surface polaron, further experiments are necessary.
Resonant photoelectron diffraction does not rule out the formation of surface localized charge carriers, even at low \cvo\cite{Kruger2008}.
SPM measurements in the presence of CO adsorbates confirm the formation of the [1$\bar{1}$2]-aligned \VO-polaron complex~\cite{reticcioli_interplay_2019}, while STM measurements probing the filled states on the clean surface do show some disparity in comparison to simulated STM\cite{reticcioli_formation_2018}.
The reasons for this discrepancy are manifold, ranging from temperature-induced effects~\cite{Reticcioli2022}, to the electric field of the tip.

We expect our methodology to be applicable to any other polaronic system, even including multiple defects as sources of polarons, such as the perovskite SrTiO$_3$(001) surface \cite{sokolovic2019incipient} exhibiting Sr adatom/vacancy and often doped by Nb atoms~\cite{ellinger_small_2023}.
Moreover, this methodology could be used to study the spatial distribution of defects (\eg\ subsurface, bulk) that are not directly accessible by the experiments, such as interstitial titanium in rutile.
Additionally, the stochastic optimization model could be further improved by considering realistic anisotropic diffusion probabilities along certain directions.
This could be achieved by explicitly computing hopping and diffusion barriers, and incorporating these barriers into the annealing simulations.

\section{Methods}
\subsection{DFT Modeling} \label{methods:DFT}
    We performed DFT+\textit{U} calculations using VASP~\cite{kresse_efficiency_1996, kresse_ultrasoft_1999, blochl_projector_1994} on the rutile \TO(110) surface.
    We used standard projector augmented wave pseudopotentials for Ti (treating $d$- and $s$-orbitals as valence) and soft Oxygen pseudopotentials.
    We adopted a Hubbard \textit{U}$=3.9$~eV on the $d$ orbitals of Ti atoms~\cite{Dudarev1998, birschitzky_machine_2022}.
    The sampling of the reciprocal space included the $\Gamma$-point and the plane-wave energy cutoff was set to 400~eV.
    
    The surfaces were modeled using 5-layer-thick slabs (where the two bottom stochiometric layers were fixed at their bulk position) with lateral supercell sizes of 6$\times$4 and 12$\times$2.
    To partially account for the role of thermal effects in the stabilization of the \VO\ patterns during the annealing treatment in the experiments, we used an expanded [001] lattice vector.
    Specifically, the low T lattice constant of 2.953~\AA~\cite{Diebold2003} was expanded to 2.968~\AA\ (high T corresponding to 500-600K) in accordance with thermal expansion coefficient measurements\cite{kirby_thermal_1967, hummer_thermal_2007}.
    This strain of $+0.5$\% is well below the crossover point of $+3$\%, where surface polaron formation is favored over subsurface polaron formation\cite{Reticcioli2022}.
    
    Within the supercells, we removed 4 surface bridging oxygen atoms (in random positions) from every slab, obtaining a \cvo\ of approximately $17$\%.
    To assess non-polaronic solutions we performed spin unpolarized DFT, constraining the excess electrons in spatially delocalized states at the bottom of the conduction band.    
    To model the polaronic structure, we followed a three-step procedure:
    Initially, we removed bridging oxygen atoms from a pristine structure to generate a specific oxygen vacancy pattern.
    This structure was relaxed while all excess charge carriers were kept delocalized by employing a spin un-polarized relaxation.
    After retrieving the structural properties of the oxygen vacancy configuration, we introduced polaronic distortions at selected sites via occupation matrix control~\cite{allen_occupation_2014}, using distinct occupation matrices for \PolSi\ and \PolSo\ sites~\cite{birschitzky_machine_2022}.
    Finally, we performed an unconstrained relaxation starting from the structures and wave functions determined in the previous step.

\subsection{ML Model Training and Optimization of Defect Configurations}
    \label{methods:ML}

    The machine learning model is implemented in the framework of JAX~\cite{bradbury_jax_2018}.
    Here, we describe the model optimization based on the study of configurations in the 6$\times$4 supercell.
    We optimized the machine learning model using stochastic gradient descent and backpropagation on an augmented dataset, including all symmetrically equivalent representations, of the training defect configurations. 
    We randomly split this dataset into 80\% training data and 20\% validation data and optimized the model parameters by minimizing the mean squared error of the energy prediction of the training data via backpropagation. 
    Before training, energies, as well as the descriptors, were rescaled to $[0,1]$, by min-max scaling. 
    Using an early stopping mechanism, the best model was selected based on the lowest validation dataset error within the optimization procedure. 
    The mean squared error during training as well as a scatter plot of DFT and ML energies are displayed in Supplementary Fig.~S1 and S2.
    
    To ensure sufficient accuracy when using the model in the case of exploration, we applied an active learning procedure as depicted in Supplementary Fig.~S3. 
    Here, we performed an iterative training-testing loop to further improve the reliability, data efficiency, and scope of the proposed model. 
    Since our main interest lies in the determination of low-energy polaron-defect complexes, our model was used for the optimization of defect configurations in various cases. 
    We searched for global optima of configurations by allowing all defects to diffuse during the optimization. 
    Local minima of fixed polaron layer densities were added by restricting polaron movement to intra-layer hopping. 
    Also, local minima of cases where the \VO-configuration was fixed and only polarons were relaxed, were explored. 
    Within these three exploration cases, we extracted and confirmed the most stable configurations by performing a comparative DFT calculation of the proposed polaron configurations.

    The optimization of configurations is performed via simulated annealing~\cite{kirkpatrick_optimization_1983}, where the temperature variable in the Metropolis criterion was set to 1000K (similar to the annealing temperatures in the sample preparation).
    Even though the diffusion processes of the respective defects during the optimization are physically motivated, they do not necessarily represent the physical process of the formation of observed defect patterns.
    Defect transport mechanisms such as inter-row hopping of oxygen vacancies have not been reported~\cite{zhang_imaging_2007} but may improve optimization efficiency or more efficiently overcome energy barriers. 
    Discrepancies between polaron and vacancy hopping rates were also ignored, which potentially affects the final outcome of the optimization.
    Similar effects were observed for the specific temperature or temperature ramp employed in the simulated annealing.

\subsection{Experimental Setup}
\label{methods:exp}
    SPM was performed using STM in an ultrahigh vacuum (UHV) chamber with a base pressure below 2$\times10^{-11}$\,mbar; the whole chamber, equipped with an Omicron qPlus low-temperature head, was suspended using 36 bungee cords for efficient vibration damping~\cite{schmid2019device}. 
    Stiff qPlus sensors~\cite{giessibl2019qplus} ($k=1800$\,N$\cdot$m$^{-1}$, $Q$=5000--30000, $f_0$\,$\in$\,[25-45]\,kHz) with a  a sharp W tip~\cite{setvin2012ultrasharp} were used to collect the tunneling current ($I_\mathrm{t}$) and the frequency shift ($\Delta f$) signals; deflection detection was achieved using a cryogenic preamplifier in vacuum~\cite{huber2017low}. 
    W tips were treated at a Cu(110) surface decorated with a sharp, conductive Cu pyramid at the apex, and were subsequently applied for imaging the rutile TiO$_2$(110) surface. 
    Tip sharpness was indicated by the low frequency shifts ($\Delta f$\,$\in$\,$[0,-1]$\,Hz) recorded during STM imaging of a Cu(110) test sample.

    Sample preparation was performed in a separate UHV chamber (connected to the measurement chamber $via$ a gate valve for \textit{in-situ} transfer) with a base pressure below  1$\times$10$^{-10}$\,mbar. 
    Surfaces were cleaned by cycles of sputtering and UHV annealing that consequently reduced the samples and introduced V$_\mathrm{O}$s to the surface. A typical cleaning cycle consisted of sputtering with 1.5\,keV Ar$^+$ ions for 10\,min with an ion current of 1\,$\mu$A$\cdot$\,cm$^{-2}$, and subsequently annealing the sputtered surfaces in UHV up to 700\,$^\circ$C. 
    Before each measurement, 3--5 cleaning cycles were performed. 
    The over-reduction of the surface was occasionally remedied by annealing the sample to 750\,$^\circ$C in 5$\times10^{-7}$\,mbar of O$_2$ shower for 10\,min. 
    When the reduction level was too high, the rutile TiO$_2$ samples were re-oxidized $ex-situ$ at 800\,$^\circ$C in O$_2$ flow and reintroduced to UHV for cleaning.

    Figure \ref{fig:STM}b displays the $z$-channel of a feedback-controlled unoccupied-states STM image taken at a sample temperature of 14\,K; Imaging parameters: sample bias $V_\mathrm{S}$ = +0.9\,V, grounded tip, tunneling current set-point $I_\mathrm{t}$=20\,pA, oscillation amplitude $A$=500\,pm.

    The contrast in Fig.\,1b corresponds to a typical unoccupied-states STM imaging contrast over a reduced rutile TiO$_2$(110) surface, which is dominated by electronic rather than geometric considerations: 1\,eV above the Fermi level the conduction band consists of Ti\,$3d$ states and defect V$_\mathrm{O}$ states, while the O states constitute the valence band \cite{diebold1996evidence}. Therefore, the highest probability of electron tunneling from the tip to the surface is above the Ti$_\mathrm{5c}$ rows and V$_\mathrm{O}$s – they appear bright under these STM conditions. On the other hand, the tunneling is less likely above the O$_\mathrm{br}$ rows and they appear dark even though they geometrically protrude highest from the surface. Note that in Fig.\,1b individual Ti$_\mathrm{5c}$ atoms can be recognized as spheres forming a row along the $[001]$ direction, while V$_\mathrm{O}$s are recognized as isolated, bright spheres.

\section{Data availability}
    The data presented in this article is available with the accompanying code upon publication or from the corresponding author upon request.

\section{Code availability}
    A minimal example of the code used to produce the presented results is available in a Github repository upon publication.
    
\section{Author Contributions}
    VB implemented the ML model and performed calculations together with MP and MR.
    VB wrote the first draft under the supervision of CF and MR.
    CF and MR conceptualized and supervised the work.
    IS and MS performed experiments under the supervision of UD.
    All authors contributed to reviewing and editing the final draft.

\section{Competing Interests}
    The authors declare no competing interests.

\begin{acknowledgement}

    This work was supported by the Austrian Science Fund (FWF) project SFB-F81 project TACO. The computational results have been achieved using the Vienna Scientific Cluster (VSC). 
    CF acknowledges the NextGenerationEU-Piano Nazionale Resistenza e Resilienza (PNRR) CN-HPC grant no. (CUP) J33C22001170001, SPOKE 6 - Multiscale Modelling \& Engineering.
    KP acknowledges the NRDIO-Hungary grant no. FK124100 and a Bolyai Fellowship of the Hungarian Academy of Sciences.

\end{acknowledgement}

\bibliography{bib}

\providecommand{\latin}[1]{#1}
\makeatletter
\providecommand{\doi}
  {\begingroup\let\do\@makeother\dospecials
  \catcode`\{=1 \catcode`\}=2 \doi@aux}
\providecommand{\doi@aux}[1]{\endgroup\texttt{#1}}
\makeatother
\providecommand*\mcitethebibliography{\thebibliography}
\csname @ifundefined\endcsname{endmcitethebibliography}
  {\let\endmcitethebibliography\endthebibliography}{}
\begin{mcitethebibliography}{71}
\providecommand*\natexlab[1]{#1}
\providecommand*\mciteSetBstSublistMode[1]{}
\providecommand*\mciteSetBstMaxWidthForm[2]{}
\providecommand*\mciteBstWouldAddEndPuncttrue
  {\def\EndOfBibitem{\unskip.}}
\providecommand*\mciteBstWouldAddEndPunctfalse
  {\let\EndOfBibitem\relax}
\providecommand*\mciteSetBstMidEndSepPunct[3]{}
\providecommand*\mciteSetBstSublistLabelBeginEnd[3]{}
\providecommand*\EndOfBibitem{}
\mciteSetBstSublistMode{f}
\mciteSetBstMaxWidthForm{subitem}{(\alph{mcitesubitemcount})}
\mciteSetBstSublistLabelBeginEnd
  {\mcitemaxwidthsubitemform\space}
  {\relax}
  {\relax}

\bibitem[Rousseau \latin{et~al.}(2020)Rousseau, Glezakou, and
  Selloni]{rousseau_theoretical_2020}
Rousseau,~R.; Glezakou,~V.-A.; Selloni,~A. Theoretical insights into the
  surface physics and chemistry of redox-active oxides. \emph{Nature Reviews
  Materials} \textbf{2020}, \emph{5}, 460--475\relax
\mciteBstWouldAddEndPuncttrue
\mciteSetBstMidEndSepPunct{\mcitedefaultmidpunct}
{\mcitedefaultendpunct}{\mcitedefaultseppunct}\relax
\EndOfBibitem
\bibitem[Franceschi and Diebold(2023)Franceschi, and Diebold]{Franceschi2022}
Franceschi,~G.; Diebold,~U. \emph{Encyclopedia of Materials: Electronics};
  2023; pp 501--511\relax
\mciteBstWouldAddEndPuncttrue
\mciteSetBstMidEndSepPunct{\mcitedefaultmidpunct}
{\mcitedefaultendpunct}{\mcitedefaultseppunct}\relax
\EndOfBibitem
\bibitem[Jupille and Thornton(2015)Jupille, and Thornton]{Jupille2015book}
Jupille,~J.; Thornton,~G. \emph{Defects at Oxide Surfaces}; Springer Series in
  Surface Sciences; 2015; Vol.~58; pp 327--349\relax
\mciteBstWouldAddEndPuncttrue
\mciteSetBstMidEndSepPunct{\mcitedefaultmidpunct}
{\mcitedefaultendpunct}{\mcitedefaultseppunct}\relax
\EndOfBibitem
\bibitem[Strand and Shluger(2023)Strand, and Shluger]{strand_structure_nodate}
Strand,~J.; Shluger,~A.~L. On the {Structure} of {Oxygen} {Deficient}
  {Amorphous} {Oxide} {Films}. \emph{Advanced Science} \textbf{2023},
  2306243\relax
\mciteBstWouldAddEndPuncttrue
\mciteSetBstMidEndSepPunct{\mcitedefaultmidpunct}
{\mcitedefaultendpunct}{\mcitedefaultseppunct}\relax
\EndOfBibitem
\bibitem[Franchini \latin{et~al.}(2021)Franchini, Reticcioli, Setvin, and
  Diebold]{Franchini2021}
Franchini,~C.; Reticcioli,~M.; Setvin,~M.; Diebold,~U. {Polarons in materials}.
  \emph{Nature Reviews Materials} \textbf{2021}, \emph{6}, 560--586\relax
\mciteBstWouldAddEndPuncttrue
\mciteSetBstMidEndSepPunct{\mcitedefaultmidpunct}
{\mcitedefaultendpunct}{\mcitedefaultseppunct}\relax
\EndOfBibitem
\bibitem[Emin(2013)]{Emin2013}
Emin,~D. \emph{{Polarons}}; Cambridge University Press, 2013\relax
\mciteBstWouldAddEndPuncttrue
\mciteSetBstMidEndSepPunct{\mcitedefaultmidpunct}
{\mcitedefaultendpunct}{\mcitedefaultseppunct}\relax
\EndOfBibitem
\bibitem[Alexandrov and Devreese(2010)Alexandrov, and Devreese]{Alexandrov2010}
Alexandrov,~A.~S.; Devreese,~J.~T. \emph{Advances in Polaron Physics}; Springer
  Series in Solid-State Sciences; 2010; Vol. 159; p 171\relax
\mciteBstWouldAddEndPuncttrue
\mciteSetBstMidEndSepPunct{\mcitedefaultmidpunct}
{\mcitedefaultendpunct}{\mcitedefaultseppunct}\relax
\EndOfBibitem
\bibitem[Stoneham \latin{et~al.}(2007)Stoneham, Gavartin, Shluger, Kimmel,
  Mũoz~Ramo, R{\o}nnow, Aeppli, and Renner]{Stoneham2007}
Stoneham,~A.~M.; Gavartin,~J.; Shluger,~A.~L.; Kimmel,~A.~V.; Mũoz~Ramo,~D.;
  R{\o}nnow,~H.~M.; Aeppli,~G.; Renner,~C. {Trapping, self-trapping and the
  polaron family}. \emph{Journal of Physics Condensed Matter} \textbf{2007},
  \emph{19}, 255208\relax
\mciteBstWouldAddEndPuncttrue
\mciteSetBstMidEndSepPunct{\mcitedefaultmidpunct}
{\mcitedefaultendpunct}{\mcitedefaultseppunct}\relax
\EndOfBibitem
\bibitem[Pastor \latin{et~al.}(2022)Pastor, Sachs, Selim, Durrant, Bakulin, and
  Walsh]{pastor_electronic_2022}
Pastor,~E.; Sachs,~M.; Selim,~S.; Durrant,~J.~R.; Bakulin,~A.~A.; Walsh,~A.
  Electronic defects in metal oxide photocatalysts. \emph{Nature Reviews
  Materials} \textbf{2022}, \emph{7}, 503--521\relax
\mciteBstWouldAddEndPuncttrue
\mciteSetBstMidEndSepPunct{\mcitedefaultmidpunct}
{\mcitedefaultendpunct}{\mcitedefaultseppunct}\relax
\EndOfBibitem
\bibitem[Kick \latin{et~al.}(2020)Kick, Grosu, Schuderer, Scheurer, and
  Oberhofer]{kick_mobile_2020}
Kick,~M.; Grosu,~C.; Schuderer,~M.; Scheurer,~C.; Oberhofer,~H. Mobile {Small}
  {Polarons} {Qualitatively} {Explain} {Conductivity} in {Lithium} {Titanium}
  {Oxide} {Battery} {Electrodes}. \emph{The Journal of Physical Chemistry
  Letters} \textbf{2020}, \emph{11}, 2535--2540\relax
\mciteBstWouldAddEndPuncttrue
\mciteSetBstMidEndSepPunct{\mcitedefaultmidpunct}
{\mcitedefaultendpunct}{\mcitedefaultseppunct}\relax
\EndOfBibitem
\bibitem[Chen \latin{et~al.}(2023)Chen, Grieder, Smart, Mayford, McNair,
  Pinongcos, Eisenberg, Bridges, Li, and Ping]{Chen2023}
Chen,~M.; Grieder,~A.~C.; Smart,~T.~J.; Mayford,~K.; McNair,~S.; Pinongcos,~A.;
  Eisenberg,~S.; Bridges,~F.; Li,~Y.; Ping,~Y. The impacts of dopants on the
  small polaron mobility and conductivity in hematite – the role of disorder.
  \emph{Nanoscale} \textbf{2023}, \emph{15}, 1619--1628\relax
\mciteBstWouldAddEndPuncttrue
\mciteSetBstMidEndSepPunct{\mcitedefaultmidpunct}
{\mcitedefaultendpunct}{\mcitedefaultseppunct}\relax
\EndOfBibitem
\bibitem[Smart and Ping(2017)Smart, and Ping]{Smart2017}
Smart,~T.~J.; Ping,~Y. {Effect of defects on the small polaron formation and
  transport properties of hematite from first-principles calculations}.
  \emph{Journal of Physics Condensed Matter} \textbf{2017}, \emph{29},
  394006\relax
\mciteBstWouldAddEndPuncttrue
\mciteSetBstMidEndSepPunct{\mcitedefaultmidpunct}
{\mcitedefaultendpunct}{\mcitedefaultseppunct}\relax
\EndOfBibitem
\bibitem[Cheng \latin{et~al.}(2022)Cheng, Zhu, Fang, Long, and
  Prezhdo]{cheng_co_2022}
Cheng,~C.; Zhu,~Y.; Fang,~W.-H.; Long,~R.; Prezhdo,~O.~V. {CO} {Adsorbate}
  {Promotes} {Polaron} {Photoactivity} on the {Reduced} {Rutile} {TiO$_2$}(110)
  {Surface}. \emph{JACS Au} \textbf{2022}, \emph{2}, 234--245\relax
\mciteBstWouldAddEndPuncttrue
\mciteSetBstMidEndSepPunct{\mcitedefaultmidpunct}
{\mcitedefaultendpunct}{\mcitedefaultseppunct}\relax
\EndOfBibitem
\bibitem[Cheng \latin{et~al.}(2022)Cheng, Zhu, Zhou, Long, and
  Fang]{cheng_photoinduced_2022}
Cheng,~C.; Zhu,~Y.; Zhou,~Z.; Long,~R.; Fang,~W.-H. Photoinduced small electron
  polarons generation and recombination in hematite. \emph{npj Computational
  Materials} \textbf{2022}, \emph{8}, 1--8\relax
\mciteBstWouldAddEndPuncttrue
\mciteSetBstMidEndSepPunct{\mcitedefaultmidpunct}
{\mcitedefaultendpunct}{\mcitedefaultseppunct}\relax
\EndOfBibitem
\bibitem[Sokolovi{\'{c}} \latin{et~al.}(2020)Sokolovi{\'{c}}, Reticcioli,
  {\v{C}}alkovsk{\'{y}}, Wagner, Schmid, Franchini, Diebold, and
  Setv{\'{i}}n]{Sokolovic2020}
Sokolovi{\'{c}},~I.; Reticcioli,~M.; {\v{C}}alkovsk{\'{y}},~M.; Wagner,~M.;
  Schmid,~M.; Franchini,~C.; Diebold,~U.; Setv{\'{i}}n,~M. {Resolving the
  adsorption of molecular O$_2$ on the rutile TiO$_2$(110) surface by
  noncontact atomic force microscopy}. \emph{Proceedings of the National
  Academy of Sciences of the United States of America} \textbf{2020},
  \emph{117}, 14827--14837\relax
\mciteBstWouldAddEndPuncttrue
\mciteSetBstMidEndSepPunct{\mcitedefaultmidpunct}
{\mcitedefaultendpunct}{\mcitedefaultseppunct}\relax
\EndOfBibitem
\bibitem[Tanner \latin{et~al.}(2021)Tanner, Wen, Ontaneda, Zhang, Grau-Crespo,
  Fielding, Selloni, and Thornton]{tanner_polaron-adsorbate_2021}
Tanner,~A.~J.; Wen,~B.; Ontaneda,~J.; Zhang,~Y.; Grau-Crespo,~R.;
  Fielding,~H.~H.; Selloni,~A.; Thornton,~G. Polaron-{Adsorbate} {Coupling} at
  the {TiO$_2$}(110)-{Carboxylate} {Interface}. \emph{The Journal of Physical
  Chemistry Letters} \textbf{2021}, \emph{12}, 3571--3576\relax
\mciteBstWouldAddEndPuncttrue
\mciteSetBstMidEndSepPunct{\mcitedefaultmidpunct}
{\mcitedefaultendpunct}{\mcitedefaultseppunct}\relax
\EndOfBibitem
\bibitem[Yim \latin{et~al.}(2018)Yim, Chen, Zhang, Shaw, Pang, Grinter, Bluhm,
  Salmeron, Muryn, Michaelides, and Thornton]{yim_visualization_2018}
Yim,~C.~M.; Chen,~J.; Zhang,~Y.; Shaw,~B.-J.; Pang,~C.~L.; Grinter,~D.~C.;
  Bluhm,~H.; Salmeron,~M.; Muryn,~C.~A.; Michaelides,~A.; Thornton,~G.
  Visualization of {Water}-{Induced} {Surface} {Segregation} of {Polarons} on
  {Rutile} {TiO$_2$}(110). \emph{The Journal of Physical Chemistry Letters}
  \textbf{2018}, \emph{9}, 4865--4871\relax
\mciteBstWouldAddEndPuncttrue
\mciteSetBstMidEndSepPunct{\mcitedefaultmidpunct}
{\mcitedefaultendpunct}{\mcitedefaultseppunct}\relax
\EndOfBibitem
\bibitem[Cheng \latin{et~al.}(2023)Cheng, Zhou, and Long]{run_photo_2023}
Cheng,~C.; Zhou,~Z.; Long,~R. Time-Domain View of Polaron Dynamics in Metal
  Oxide Photocatalysts. \emph{The Journal of Physical Chemistry Letters}
  \textbf{2023}, \emph{14}, 10988--10998\relax
\mciteBstWouldAddEndPuncttrue
\mciteSetBstMidEndSepPunct{\mcitedefaultmidpunct}
{\mcitedefaultendpunct}{\mcitedefaultseppunct}\relax
\EndOfBibitem
\bibitem[Ren \latin{et~al.}(2023)Ren, Shi, Feng, Xu, and Hao]{ren_recent_2023}
Ren,~Z.; Shi,~Z.; Feng,~H.; Xu,~Z.; Hao,~W. Recent {Progresses} of {Polarons}:
  {Fundamentals} and {Roles} in {Photocatalysis} and {Photoelectrocatalysis}.
  \emph{Advanced Science} \textbf{2023}, 2305139\relax
\mciteBstWouldAddEndPuncttrue
\mciteSetBstMidEndSepPunct{\mcitedefaultmidpunct}
{\mcitedefaultendpunct}{\mcitedefaultseppunct}\relax
\EndOfBibitem
\bibitem[Dohnálek \latin{et~al.}(2010)Dohnálek, Lyubinetsky, and
  Rousseau]{dohnalek_thermally-driven_2010}
Dohnálek,~Z.; Lyubinetsky,~I.; Rousseau,~R. Thermally-driven processes on
  rutile {TiO$_2$}(110)-(1×1): {A} direct view at the atomic scale.
  \emph{Progress in Surface Science} \textbf{2010}, \emph{85}, 161--205\relax
\mciteBstWouldAddEndPuncttrue
\mciteSetBstMidEndSepPunct{\mcitedefaultmidpunct}
{\mcitedefaultendpunct}{\mcitedefaultseppunct}\relax
\EndOfBibitem
\bibitem[Tanner and Thornton(2022)Tanner, and Thornton]{tanner_tio2_2022}
Tanner,~A.~J.; Thornton,~G. {TiO$_2$} {Polarons} in the {Time} {Domain}:
  {Implications} for {Photocatalysis}. \emph{The Journal of Physical Chemistry
  Letters} \textbf{2022}, \emph{13}, 559--566\relax
\mciteBstWouldAddEndPuncttrue
\mciteSetBstMidEndSepPunct{\mcitedefaultmidpunct}
{\mcitedefaultendpunct}{\mcitedefaultseppunct}\relax
\EndOfBibitem
\bibitem[Sombut \latin{et~al.}(2022)Sombut, Puntscher, Atzmueller, Jakub,
  Reticcioli, Meier, Parkinson, and Franchini]{sombut_role_2022}
Sombut,~P.; Puntscher,~L.; Atzmueller,~M.; Jakub,~Z.; Reticcioli,~M.;
  Meier,~M.; Parkinson,~G.~S.; Franchini,~C. {Role of Polarons in Single-Atom
  Catalysts: Case Study of Me1 [Au1, Pt1, and Rh1] on TiO$_2$(110)}.
  \emph{Topics in Catalysis} \textbf{2022}, \emph{2}, 1--16\relax
\mciteBstWouldAddEndPuncttrue
\mciteSetBstMidEndSepPunct{\mcitedefaultmidpunct}
{\mcitedefaultendpunct}{\mcitedefaultseppunct}\relax
\EndOfBibitem
\bibitem[Geiger and López(2022)Geiger, and López]{geiger_coupling_2022}
Geiger,~J.; López,~N. Coupling {Metal} and {Support} {Redox} {Terms} in
  {Single}-{Atom} {Catalysts}. \emph{The Journal of Physical Chemistry C}
  \textbf{2022}, \emph{126}, 13698--13704\relax
\mciteBstWouldAddEndPuncttrue
\mciteSetBstMidEndSepPunct{\mcitedefaultmidpunct}
{\mcitedefaultendpunct}{\mcitedefaultseppunct}\relax
\EndOfBibitem
\bibitem[Geiger \latin{et~al.}(2022)Geiger, Sabadell-Rendón, Daelman, and
  López]{geiger_data-driven_2022}
Geiger,~J.; Sabadell-Rendón,~A.; Daelman,~N.; López,~N. Data-driven models
  for ground and excited states for {Single} {Atoms} on {Ceria}. \emph{npj
  Computational Materials} \textbf{2022}, \emph{8}, 1--8\relax
\mciteBstWouldAddEndPuncttrue
\mciteSetBstMidEndSepPunct{\mcitedefaultmidpunct}
{\mcitedefaultendpunct}{\mcitedefaultseppunct}\relax
\EndOfBibitem
\bibitem[Cao \latin{et~al.}(2017)Cao, Yu, Qi, Huang, Wang, Xu, Hu, and
  Yan]{cao_scenarios_2017}
Cao,~Y.; Yu,~M.; Qi,~S.; Huang,~S.; Wang,~T.; Xu,~M.; Hu,~S.; Yan,~S. Scenarios
  of polaron-involved molecular adsorption on reduced {TiO$_2$}(110) surfaces.
  \emph{Scientific Reports} \textbf{2017}, \emph{7}, 6148\relax
\mciteBstWouldAddEndPuncttrue
\mciteSetBstMidEndSepPunct{\mcitedefaultmidpunct}
{\mcitedefaultendpunct}{\mcitedefaultseppunct}\relax
\EndOfBibitem
\bibitem[Reticcioli \latin{et~al.}(2019)Reticcioli, Sokolović, Schmid,
  Diebold, Setvin, and Franchini]{reticcioli_interplay_2019}
Reticcioli,~M.; Sokolović,~I.; Schmid,~M.; Diebold,~U.; Setvin,~M.;
  Franchini,~C. Interplay between {Adsorbates} and {Polarons}: {CO} on {Rutile}
  {TiO} $_{\textrm{2}}$(110). \emph{Physical Review Letters} \textbf{2019},
  \emph{122}, 016805\relax
\mciteBstWouldAddEndPuncttrue
\mciteSetBstMidEndSepPunct{\mcitedefaultmidpunct}
{\mcitedefaultendpunct}{\mcitedefaultseppunct}\relax
\EndOfBibitem
\bibitem[Birschitzky \latin{et~al.}(2022)Birschitzky, Ellinger, Diebold,
  Reticcioli, and Franchini]{birschitzky_machine_2022}
Birschitzky,~V.~C.; Ellinger,~F.; Diebold,~U.; Reticcioli,~M.; Franchini,~C.
  Machine learning for exploring small polaron configurational space. \emph{npj
  Computational Materials} \textbf{2022}, \emph{8}, 1--9\relax
\mciteBstWouldAddEndPuncttrue
\mciteSetBstMidEndSepPunct{\mcitedefaultmidpunct}
{\mcitedefaultendpunct}{\mcitedefaultseppunct}\relax
\EndOfBibitem
\bibitem[Zhang \latin{et~al.}(2019)Zhang, Han, Murgida, Ganduglia-Pirovano, and
  Gao]{Zhang2019}
Zhang,~D.; Han,~Z.~K.; Murgida,~G.~E.; Ganduglia-Pirovano,~M.~V.; Gao,~Y.
  {Oxygen-Vacancy Dynamics and Entanglement with Polaron Hopping at the Reduced
  CeO$_2$ (111) Surface}. \emph{Physical Review Letters} \textbf{2019},
  \emph{122}, 096101\relax
\mciteBstWouldAddEndPuncttrue
\mciteSetBstMidEndSepPunct{\mcitedefaultmidpunct}
{\mcitedefaultendpunct}{\mcitedefaultseppunct}\relax
\EndOfBibitem
\bibitem[Ellinger \latin{et~al.}(2023)Ellinger, Shafiq, Ahmad, Reticcioli, and
  Franchini]{ellinger_small_2023}
Ellinger,~F.; Shafiq,~M.; Ahmad,~I.; Reticcioli,~M.; Franchini,~C. {Small
  Polaron Formation on the Nb-doped SrTiO$_3$(001) Surface}. \emph{Physical
  Review Materials} \textbf{2023}, \emph{7}, 064602\relax
\mciteBstWouldAddEndPuncttrue
\mciteSetBstMidEndSepPunct{\mcitedefaultmidpunct}
{\mcitedefaultendpunct}{\mcitedefaultseppunct}\relax
\EndOfBibitem
\bibitem[Österbacka \latin{et~al.}(2022)Österbacka, Ambrosio, and
  Wiktor]{osterbacka_charge_2022}
Österbacka,~N.; Ambrosio,~F.; Wiktor,~J. Charge {Localization} in {Defective}
  {BiVO}$_{\textrm{4}}$. \emph{The Journal of Physical Chemistry C}
  \textbf{2022}, \emph{126}, 2960--2970\relax
\mciteBstWouldAddEndPuncttrue
\mciteSetBstMidEndSepPunct{\mcitedefaultmidpunct}
{\mcitedefaultendpunct}{\mcitedefaultseppunct}\relax
\EndOfBibitem
\bibitem[Sun \latin{et~al.}(2017)Sun, Huang, Wang, and Janotti]{Sun2017}
Sun,~L.; Huang,~X.; Wang,~L.; Janotti,~A. {Disentangling the role of small
  polarons and oxygen vacancies in CeO$_2$}. \emph{Physical Review B}
  \textbf{2017}, \emph{95}, 245101\relax
\mciteBstWouldAddEndPuncttrue
\mciteSetBstMidEndSepPunct{\mcitedefaultmidpunct}
{\mcitedefaultendpunct}{\mcitedefaultseppunct}\relax
\EndOfBibitem
\bibitem[Reticcioli \latin{et~al.}(2019)Reticcioli, Diebold, Kresse, and
  Franchini]{Reticcioli2019b}
Reticcioli,~M.; Diebold,~U.; Kresse,~G.; Franchini,~C. \emph{Handbook of
  Materials Modeling}; Springer International Publishing, 2019; pp 1--39\relax
\mciteBstWouldAddEndPuncttrue
\mciteSetBstMidEndSepPunct{\mcitedefaultmidpunct}
{\mcitedefaultendpunct}{\mcitedefaultseppunct}\relax
\EndOfBibitem
\bibitem[Pham and Deskins(2020)Pham, and Deskins]{Pham2020}
Pham,~T.~D.; Deskins,~N.~A. {Efficient Method for Modeling Polarons Using
  Electronic Structure Methods}. \emph{Journal of Chemical Theory and
  Computation} \textbf{2020}, \emph{16}, 5264--5278\relax
\mciteBstWouldAddEndPuncttrue
\mciteSetBstMidEndSepPunct{\mcitedefaultmidpunct}
{\mcitedefaultendpunct}{\mcitedefaultseppunct}\relax
\EndOfBibitem
\bibitem[Reticcioli \latin{et~al.}(2017)Reticcioli, Setvin, Hao, Flauger,
  Kresse, Schmid, Diebold, and Franchini]{Reticcioli2017d}
Reticcioli,~M.; Setvin,~M.; Hao,~X.; Flauger,~P.; Kresse,~G.; Schmid,~M.;
  Diebold,~U.; Franchini,~C. {Polaron-driven surface reconstructions}.
  \emph{Physical Review X} \textbf{2017}, \emph{7}, 031053\relax
\mciteBstWouldAddEndPuncttrue
\mciteSetBstMidEndSepPunct{\mcitedefaultmidpunct}
{\mcitedefaultendpunct}{\mcitedefaultseppunct}\relax
\EndOfBibitem
\bibitem[Kowalski \latin{et~al.}(2010)Kowalski, Camellone, Nair, Meyer, and
  Marx]{Kowalski2010}
Kowalski,~P.~M.; Camellone,~M.~F.; Nair,~N.~N.; Meyer,~B.; Marx,~D. {Charge
  localization dynamics induced by oxygen vacancies on the TiO$_2$(110)
  surface}. \emph{Physical Review Letters} \textbf{2010}, \emph{105},
  146405\relax
\mciteBstWouldAddEndPuncttrue
\mciteSetBstMidEndSepPunct{\mcitedefaultmidpunct}
{\mcitedefaultendpunct}{\mcitedefaultseppunct}\relax
\EndOfBibitem
\bibitem[Han \latin{et~al.}(2018)Han, Yang, Zhu, Ganduglia-Pirovano, and
  Gao]{Han2018}
Han,~Z.~K.; Yang,~Y.~Z.; Zhu,~B.; Ganduglia-Pirovano,~M.~V.; Gao,~Y.
  {Unraveling the oxygen vacancy structures at the reduced CeO$_2$(111)
  surface}. \emph{Physical Review Materials} \textbf{2018}, \emph{2},
  035802\relax
\mciteBstWouldAddEndPuncttrue
\mciteSetBstMidEndSepPunct{\mcitedefaultmidpunct}
{\mcitedefaultendpunct}{\mcitedefaultseppunct}\relax
\EndOfBibitem
\bibitem[Reticcioli \latin{et~al.}(2018)Reticcioli, Setvin, Schmid, Diebold,
  and Franchini]{reticcioli_formation_2018}
Reticcioli,~M.; Setvin,~M.; Schmid,~M.; Diebold,~U.; Franchini,~C. {Formation
  and dynamics of small polarons on the rutile TiO$_2$(110) surface}.
  \emph{Physical Review B} \textbf{2018}, \emph{98}, 045306\relax
\mciteBstWouldAddEndPuncttrue
\mciteSetBstMidEndSepPunct{\mcitedefaultmidpunct}
{\mcitedefaultendpunct}{\mcitedefaultseppunct}\relax
\EndOfBibitem
\bibitem[Diebold(2003)]{Diebold2003}
Diebold,~U. {The surface science of titanium dioxide}. \emph{Surface Science
  Reports} \textbf{2003}, \emph{48}, 53--229\relax
\mciteBstWouldAddEndPuncttrue
\mciteSetBstMidEndSepPunct{\mcitedefaultmidpunct}
{\mcitedefaultendpunct}{\mcitedefaultseppunct}\relax
\EndOfBibitem
\bibitem[Setvin \latin{et~al.}(2014)Setvin, Franchini, Hao, Schmid, Janotti,
  Kaltak, Van De~Walle, Kresse, and Diebold]{setvin_direct_2014}
Setvin,~M.; Franchini,~C.; Hao,~X.; Schmid,~M.; Janotti,~A.; Kaltak,~M.; Van
  De~Walle,~C.~G.; Kresse,~G.; Diebold,~U. {Direct view at excess electrons in
  TiO$_2$ rutile and anatase}. \emph{Physical Review Letters} \textbf{2014},
  \emph{113}, 086402\relax
\mciteBstWouldAddEndPuncttrue
\mciteSetBstMidEndSepPunct{\mcitedefaultmidpunct}
{\mcitedefaultendpunct}{\mcitedefaultseppunct}\relax
\EndOfBibitem
\bibitem[Onishi and Iwasawa(1994)Onishi, and Iwasawa]{Onishi1994}
Onishi,~H.; Iwasawa,~Y. {Reconstruction of TiO$_2$(110) surface: STM study with
  atomic-scale resolution}. \emph{Surface Science} \textbf{1994}, \emph{313},
  L783--L789\relax
\mciteBstWouldAddEndPuncttrue
\mciteSetBstMidEndSepPunct{\mcitedefaultmidpunct}
{\mcitedefaultendpunct}{\mcitedefaultseppunct}\relax
\EndOfBibitem
\bibitem[Li \latin{et~al.}(1999)Li, Hebenstreit, Gross, Diebold, Henderson,
  Jennison, Schultz, and Sears]{Li1999a}
Li,~M.; Hebenstreit,~W.; Gross,~L.; Diebold,~U.; Henderson,~M.~A.;
  Jennison,~D.~R.; Schultz,~P.~A.; Sears,~M.~P. {Oxygen-induced restructuring
  of the TiO$_2$(110) surface: a comprehensive study}. \emph{Surface Science}
  \textbf{1999}, \emph{437}, 173--190\relax
\mciteBstWouldAddEndPuncttrue
\mciteSetBstMidEndSepPunct{\mcitedefaultmidpunct}
{\mcitedefaultendpunct}{\mcitedefaultseppunct}\relax
\EndOfBibitem
\bibitem[Li \latin{et~al.}(2000)Li, Hebenstreit, and Diebold]{Li2000}
Li,~M.; Hebenstreit,~W.; Diebold,~U. {Morphology change of oxygen-restructured
  surfaces by UHV annealing: Formation of a low-temperature structure}.
  \emph{Physical Review B} \textbf{2000}, \emph{61}, 4926--4933\relax
\mciteBstWouldAddEndPuncttrue
\mciteSetBstMidEndSepPunct{\mcitedefaultmidpunct}
{\mcitedefaultendpunct}{\mcitedefaultseppunct}\relax
\EndOfBibitem
\bibitem[McCarty and Bartelt(2003)McCarty, and Bartelt]{McCarty2003}
McCarty,~K.~F.; Bartelt,~N.~C. {The $1\times1$/$1\times2$ phase transition of
  the TiO$_2$(110) surface-variation of transition temperature with crystal
  composition}. \emph{Surface Science} \textbf{2003}, \emph{527},
  L203--L212\relax
\mciteBstWouldAddEndPuncttrue
\mciteSetBstMidEndSepPunct{\mcitedefaultmidpunct}
{\mcitedefaultendpunct}{\mcitedefaultseppunct}\relax
\EndOfBibitem
\bibitem[Wang \latin{et~al.}(2014)Wang, Oganov, Zhu, and Zhou]{Wang2014a}
Wang,~Q.; Oganov,~A.~R.; Zhu,~Q.; Zhou,~X.~F. {New reconstructions of the (110)
  surface of rutile TiO$_2$ predicted by an evolutionary method}.
  \emph{Physical Review Letters} \textbf{2014}, \emph{113}, 266101\relax
\mciteBstWouldAddEndPuncttrue
\mciteSetBstMidEndSepPunct{\mcitedefaultmidpunct}
{\mcitedefaultendpunct}{\mcitedefaultseppunct}\relax
\EndOfBibitem
\bibitem[Mochizuki \latin{et~al.}(2016)Mochizuki, Ariga, Fukaya, Wada, Maekawa,
  Kawasuso, Shidara, Asakura, and Hyodo]{Mochizuki2016}
Mochizuki,~I.; Ariga,~H.; Fukaya,~Y.; Wada,~K.; Maekawa,~M.; Kawasuso,~A.;
  Shidara,~T.; Asakura,~K.; Hyodo,~T. {Structure determination of the
  rutile-TiO$_2$(110)-(1$\times$2) surface using total-reflection high-energy
  positron diffraction (TRHEPD)}. \emph{Physical Chemistry Chemical Physics}
  \textbf{2016}, \emph{18}, 7085--7092\relax
\mciteBstWouldAddEndPuncttrue
\mciteSetBstMidEndSepPunct{\mcitedefaultmidpunct}
{\mcitedefaultendpunct}{\mcitedefaultseppunct}\relax
\EndOfBibitem
\bibitem[Deskins \latin{et~al.}(2009)Deskins, Rousseau, and
  Dupuis]{Deskins2009}
Deskins,~N.~A.; Rousseau,~R.; Dupuis,~M. {Localized electronic states from
  surface hydroxyls and polarons in TiO$_2$(110)}. \emph{Journal of Physical
  Chemistry C} \textbf{2009}, \emph{113}, 14583--14586\relax
\mciteBstWouldAddEndPuncttrue
\mciteSetBstMidEndSepPunct{\mcitedefaultmidpunct}
{\mcitedefaultendpunct}{\mcitedefaultseppunct}\relax
\EndOfBibitem
\bibitem[LeCun \latin{et~al.}(2015)LeCun, Bengio, and Hinton]{lecun_deep_2015}
LeCun,~Y.; Bengio,~Y.; Hinton,~G. Deep learning. \emph{Nature} \textbf{2015},
  \emph{521}, 436--444\relax
\mciteBstWouldAddEndPuncttrue
\mciteSetBstMidEndSepPunct{\mcitedefaultmidpunct}
{\mcitedefaultendpunct}{\mcitedefaultseppunct}\relax
\EndOfBibitem
\bibitem[Behler(2021)]{behler_four_2021}
Behler,~J. Four {Generations} of {High}-{Dimensional} {Neural} {Network}
  {Potentials}. \emph{Chemical Reviews} \textbf{2021}, \emph{121},
  10037--10072\relax
\mciteBstWouldAddEndPuncttrue
\mciteSetBstMidEndSepPunct{\mcitedefaultmidpunct}
{\mcitedefaultendpunct}{\mcitedefaultseppunct}\relax
\EndOfBibitem
\bibitem[Kresse and Furthmüller(1996)Kresse, and
  Furthmüller]{kresse_efficiency_1996}
Kresse,~G.; Furthmüller,~J. Efficiency of ab-initio total energy calculations
  for metals and semiconductors using a plane-wave basis set.
  \emph{Computational Materials Science} \textbf{1996}, \emph{6}, 15--50\relax
\mciteBstWouldAddEndPuncttrue
\mciteSetBstMidEndSepPunct{\mcitedefaultmidpunct}
{\mcitedefaultendpunct}{\mcitedefaultseppunct}\relax
\EndOfBibitem
\bibitem[Kresse and Joubert(1999)Kresse, and Joubert]{kresse_ultrasoft_1999}
Kresse,~G.; Joubert,~D. From ultrasoft pseudopotentials to the projector
  augmented-wave method. \emph{Physical Review B} \textbf{1999}, \emph{59},
  1758--1775\relax
\mciteBstWouldAddEndPuncttrue
\mciteSetBstMidEndSepPunct{\mcitedefaultmidpunct}
{\mcitedefaultendpunct}{\mcitedefaultseppunct}\relax
\EndOfBibitem
\bibitem[Blöchl(1994)]{blochl_projector_1994}
Blöchl,~P.~E. Projector augmented-wave method. \emph{Physical Review B}
  \textbf{1994}, \emph{50}, 17953--17979\relax
\mciteBstWouldAddEndPuncttrue
\mciteSetBstMidEndSepPunct{\mcitedefaultmidpunct}
{\mcitedefaultendpunct}{\mcitedefaultseppunct}\relax
\EndOfBibitem
\bibitem[Dudarev and Botton(1998)Dudarev, and Botton]{Dudarev1998}
Dudarev,~S.; Botton,~G. {Electron-energy-loss spectra and the structural
  stability of nickel oxide: An LSDA+U study}. \emph{Physical Review B}
  \textbf{1998}, \emph{57}, 1505--1509\relax
\mciteBstWouldAddEndPuncttrue
\mciteSetBstMidEndSepPunct{\mcitedefaultmidpunct}
{\mcitedefaultendpunct}{\mcitedefaultseppunct}\relax
\EndOfBibitem
\bibitem[Allen and Watson(2014)Allen, and Watson]{allen_occupation_2014}
Allen,~J.~P.; Watson,~G.~W. Occupation matrix control of d- and f-electron
  localisations using {DFT} + {U}. \emph{Physical Chemistry Chemical Physics}
  \textbf{2014}, \emph{16}, 21016--21031\relax
\mciteBstWouldAddEndPuncttrue
\mciteSetBstMidEndSepPunct{\mcitedefaultmidpunct}
{\mcitedefaultendpunct}{\mcitedefaultseppunct}\relax
\EndOfBibitem
\bibitem[Kirkpatrick \latin{et~al.}(1983)Kirkpatrick, Gelatt, and
  Vecchi]{kirkpatrick_optimization_1983}
Kirkpatrick,~S.; Gelatt,~C.~D.; Vecchi,~M.~P. Optimization by {Simulated}
  {Annealing}. \emph{Science} \textbf{1983}, \emph{220}, 671--680\relax
\mciteBstWouldAddEndPuncttrue
\mciteSetBstMidEndSepPunct{\mcitedefaultmidpunct}
{\mcitedefaultendpunct}{\mcitedefaultseppunct}\relax
\EndOfBibitem
\bibitem[Metropolis \latin{et~al.}(1953)Metropolis, Rosenbluth, Rosenbluth,
  Teller, and Teller]{metropolis_equation_2004}
Metropolis,~N.; Rosenbluth,~A.~W.; Rosenbluth,~M.~N.; Teller,~A.~H.; Teller,~E.
  Equation of {State} {Calculations} by {Fast} {Computing} {Machines}.
  \emph{The Journal of Chemical Physics} \textbf{1953}, \emph{21},
  1087--1092\relax
\mciteBstWouldAddEndPuncttrue
\mciteSetBstMidEndSepPunct{\mcitedefaultmidpunct}
{\mcitedefaultendpunct}{\mcitedefaultseppunct}\relax
\EndOfBibitem
\bibitem[Ji \latin{et~al.}(2022)Ji, Wang, Li, Zhu, Lin, Deng, Chen, Zhang, and
  Xing]{ji_oxygen_2022}
Ji,~W.; Wang,~N.; Li,~Q.; Zhu,~H.; Lin,~K.; Deng,~J.; Chen,~J.; Zhang,~H.;
  Xing,~X. Oxygen vacancy distributions and electron localization in a
  {CeO$_2$}(100) nanocube. \emph{Inorganic Chemistry Frontiers} \textbf{2022},
  \emph{9}, 275--283\relax
\mciteBstWouldAddEndPuncttrue
\mciteSetBstMidEndSepPunct{\mcitedefaultmidpunct}
{\mcitedefaultendpunct}{\mcitedefaultseppunct}\relax
\EndOfBibitem
\bibitem[Setvin \latin{et~al.}(2015)Setvin, Buchholz, Hou, Zhang, St{\"{o}}ger,
  Hulva, Simschitz, Shi, Pavelec, Parkinson, Xu, Wang, Schmid, W{\"{o}}ll,
  Selloni, and Diebold]{Setvin2015b}
Setvin,~M. \latin{et~al.}  {A Multitechnique Study of CO Adsorption on the
  TiO$_2$ Anatase (101) Surface}. \emph{Journal of Physical Chemistry C}
  \textbf{2015}, \emph{119}, 21044--21052\relax
\mciteBstWouldAddEndPuncttrue
\mciteSetBstMidEndSepPunct{\mcitedefaultmidpunct}
{\mcitedefaultendpunct}{\mcitedefaultseppunct}\relax
\EndOfBibitem
\bibitem[Potapenko \latin{et~al.}(2014)Potapenko, Li, Kysar, and
  Osgood]{doi:10.1021/nl5024194}
Potapenko,~D.~V.; Li,~Z.; Kysar,~J.~W.; Osgood,~R.~M. Nanoscale Strain
  Engineering on the Surface of a Bulk TiO$_2$ Crystal. \emph{Nano Letters}
  \textbf{2014}, \emph{14}, 6185--6189\relax
\mciteBstWouldAddEndPuncttrue
\mciteSetBstMidEndSepPunct{\mcitedefaultmidpunct}
{\mcitedefaultendpunct}{\mcitedefaultseppunct}\relax
\EndOfBibitem
\bibitem[Kr{\"{u}}ger \latin{et~al.}(2008)Kr{\"{u}}ger, Bourgeois, Domenichini,
  Magnan, Chandesris, Le~F{\`{e}}vre, Flank, Jupille, Floreano, Cossaro,
  Verdini, and Morgante]{Kruger2008}
Kr{\"{u}}ger,~P.; Bourgeois,~S.; Domenichini,~B.; Magnan,~H.; Chandesris,~D.;
  Le~F{\`{e}}vre,~P.; Flank,~A.~M.; Jupille,~J.; Floreano,~L.; Cossaro,~A.;
  Verdini,~A.; Morgante,~A. {Defect states at the TiO$_2$(110) surface probed
  by resonant photoelectron diffraction}. \emph{Physical Review Letters}
  \textbf{2008}, \emph{100}, 055501\relax
\mciteBstWouldAddEndPuncttrue
\mciteSetBstMidEndSepPunct{\mcitedefaultmidpunct}
{\mcitedefaultendpunct}{\mcitedefaultseppunct}\relax
\EndOfBibitem
\bibitem[Reticcioli \latin{et~al.}(2022)Reticcioli, Diebold, and
  Franchini]{Reticcioli2022}
Reticcioli,~M.; Diebold,~U.; Franchini,~C. {Modeling polarons in density
  functional theory: lessons learned from TiO$_2$}. \emph{Journal of Physics:
  Condensed Matter} \textbf{2022}, \emph{34}, 204006\relax
\mciteBstWouldAddEndPuncttrue
\mciteSetBstMidEndSepPunct{\mcitedefaultmidpunct}
{\mcitedefaultendpunct}{\mcitedefaultseppunct}\relax
\EndOfBibitem
\bibitem[Sokolovi{\'c} \latin{et~al.}(2019)Sokolovi{\'c}, Schmid, Diebold, and
  Setv\'{i}n]{sokolovic2019incipient}
Sokolovi{\'c},~I.; Schmid,~M.; Diebold,~U.; Setv\'{i}n,~M. Incipient
  ferroelectricity: {A} route towards bulk-terminated {S}r{T}i{O}$_3$.
  \emph{Phys. Rev. Mater.} \textbf{2019}, \emph{3}, 034407\relax
\mciteBstWouldAddEndPuncttrue
\mciteSetBstMidEndSepPunct{\mcitedefaultmidpunct}
{\mcitedefaultendpunct}{\mcitedefaultseppunct}\relax
\EndOfBibitem
\bibitem[Kirby(1967)]{kirby_thermal_1967}
Kirby,~R.~K. Thermal {Expansion} of {Rutile} from 100 to 700 °{K}.
  \emph{Journal of Research of the National Bureau of Standards. Section A,
  Physics and Chemistry} \textbf{1967}, \emph{71A}, 363--369\relax
\mciteBstWouldAddEndPuncttrue
\mciteSetBstMidEndSepPunct{\mcitedefaultmidpunct}
{\mcitedefaultendpunct}{\mcitedefaultseppunct}\relax
\EndOfBibitem
\bibitem[Hummer \latin{et~al.}(2007)Hummer, Heaney, and
  Post]{hummer_thermal_2007}
Hummer,~D.~R.; Heaney,~P.~J.; Post,~J.~E. Thermal expansion of anatase and
  rutile between 300 and 575 {K} using synchrotron powder {X}-ray diffraction.
  \emph{Powder Diffraction} \textbf{2007}, \emph{22}, 352--357\relax
\mciteBstWouldAddEndPuncttrue
\mciteSetBstMidEndSepPunct{\mcitedefaultmidpunct}
{\mcitedefaultendpunct}{\mcitedefaultseppunct}\relax
\EndOfBibitem
\bibitem[Bradbury \latin{et~al.}(2018)Bradbury, Frostig, Hawkins, Johnson,
  Leary, Maclaurin, Necula, Paszke, VanderPlas, Wanderman-Milne, and
  Zhang]{bradbury_jax_2018}
Bradbury,~J.; Frostig,~R.; Hawkins,~P.; Johnson,~M.~J.; Leary,~C.;
  Maclaurin,~D.; Necula,~G.; Paszke,~A.; VanderPlas,~J.; Wanderman-Milne,~S.;
  Zhang,~Q. {JAX}: composable transformations of {Python}+{NumPy} programs.
  2018; \url{http://github.com/google/jax}\relax
\mciteBstWouldAddEndPuncttrue
\mciteSetBstMidEndSepPunct{\mcitedefaultmidpunct}
{\mcitedefaultendpunct}{\mcitedefaultseppunct}\relax
\EndOfBibitem
\bibitem[Zhang \latin{et~al.}(2007)Zhang, Ge, Li, Kay, White, and
  Dohnálek]{zhang_imaging_2007}
Zhang,~Z.; Ge,~Q.; Li,~S.-C.; Kay,~B.~D.; White,~J.~M.; Dohnálek,~Z. Imaging
  {Intrinsic} {Diffusion} of {Bridge}-{Bonded} {Oxygen} {Vacancies} on
  {TiO$_2$}( 110 ). \emph{Physical Review Letters} \textbf{2007}, \emph{99},
  126105\relax
\mciteBstWouldAddEndPuncttrue
\mciteSetBstMidEndSepPunct{\mcitedefaultmidpunct}
{\mcitedefaultendpunct}{\mcitedefaultseppunct}\relax
\EndOfBibitem
\bibitem[Schmid \latin{et~al.}(2019)Schmid, Setvín, and
  Diebold]{schmid2019device}
Schmid,~M.; Setvín,~M.; Diebold,~U. Device for suspending a load in a
  vibration-insulated manner. 2019\relax
\mciteBstWouldAddEndPuncttrue
\mciteSetBstMidEndSepPunct{\mcitedefaultmidpunct}
{\mcitedefaultendpunct}{\mcitedefaultseppunct}\relax
\EndOfBibitem
\bibitem[Giessibl(2019)]{giessibl2019qplus}
Giessibl,~F.~J. The q{P}lus sensor, a powerful core for the atomic force
  microscope. \emph{Rev. Sci. Instr.} \textbf{2019}, \emph{90}, 011101\relax
\mciteBstWouldAddEndPuncttrue
\mciteSetBstMidEndSepPunct{\mcitedefaultmidpunct}
{\mcitedefaultendpunct}{\mcitedefaultseppunct}\relax
\EndOfBibitem
\bibitem[Setv{\'\i}n \latin{et~al.}(2012)Setv{\'\i}n, Javorsk{\`y},
  Tur{\v{c}}inkov{\'a}, Matol{\'\i}nov{\'a}, Sobot{\'\i}k, Koc{\'a}n, and
  O{\v{s}}t’{\'a}dal]{setvin2012ultrasharp}
Setv{\'\i}n,~M.; Javorsk{\`y},~J.; Tur{\v{c}}inkov{\'a},~D.;
  Matol{\'\i}nov{\'a},~I.; Sobot{\'\i}k,~P.; Koc{\'a}n,~P.;
  O{\v{s}}t’{\'a}dal,~I. Ultrasharp tungsten tips—characterization and
  nondestructive cleaning. \emph{Ultramicroscopy} \textbf{2012}, \emph{113},
  152--157\relax
\mciteBstWouldAddEndPuncttrue
\mciteSetBstMidEndSepPunct{\mcitedefaultmidpunct}
{\mcitedefaultendpunct}{\mcitedefaultseppunct}\relax
\EndOfBibitem
\bibitem[Huber and Giessibl(2017)Huber, and Giessibl]{huber2017low}
Huber,~F.; Giessibl,~F.~J. Low noise current preamplifier for q{P}lus sensor
  deflection signal detection in atomic force microscopy at room and low
  temperatures. \emph{Rev. Sci. Instrum.} \textbf{2017}, \emph{88},
  073702\relax
\mciteBstWouldAddEndPuncttrue
\mciteSetBstMidEndSepPunct{\mcitedefaultmidpunct}
{\mcitedefaultendpunct}{\mcitedefaultseppunct}\relax
\EndOfBibitem
\bibitem[Diebold \latin{et~al.}(1996)Diebold, Anderson, Ng, and
  Vanderbilt]{diebold1996evidence}
Diebold,~U.; Anderson,~J.~F.; Ng,~K.~O.; Vanderbilt,~D. Evidence for the
  Tunneling Site on Transition-Metal Oxides: {T}i{O}$_2$(110). \emph{Phys. Rev.
  Lett.} \textbf{1996}, \emph{77}, 1322\relax
\mciteBstWouldAddEndPuncttrue
\mciteSetBstMidEndSepPunct{\mcitedefaultmidpunct}
{\mcitedefaultendpunct}{\mcitedefaultseppunct}\relax
\EndOfBibitem
\end{mcitethebibliography}
\end{document}


\newpage

\section{Machine Learning Model Training}

    \begin{figure}[!h]
        \centering
        \includegraphics[width=\textwidth]{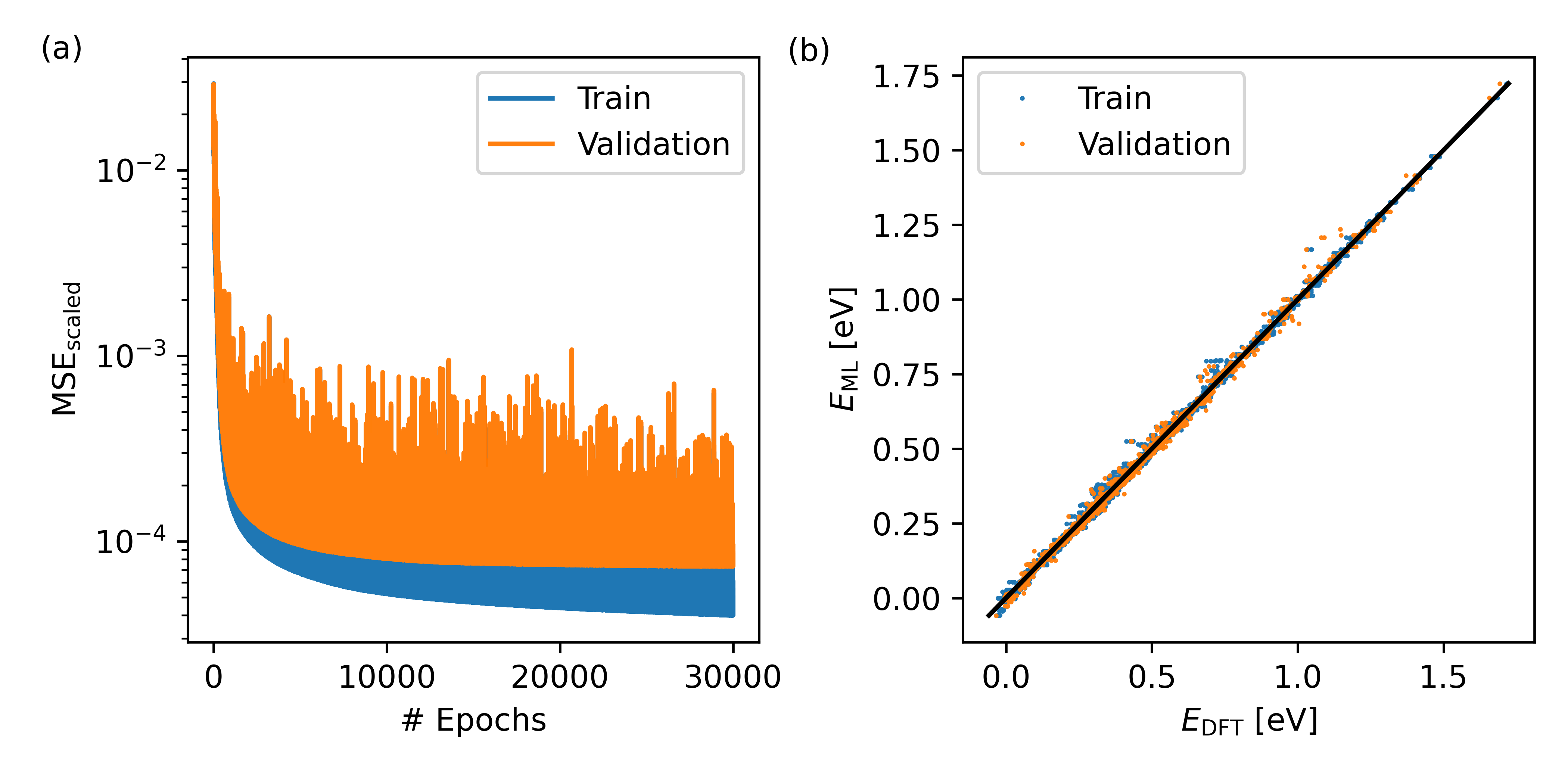}
        \caption{Training procedure of the 6$\times$4 model characterized with a (a) learning curve and a (b) scatter plot comparing DFT- and ML-based energies.}
        \label{fig_sup:train64}
    \end{figure}

    \begin{figure}[!h]
        \centering
        \includegraphics[width=\textwidth]{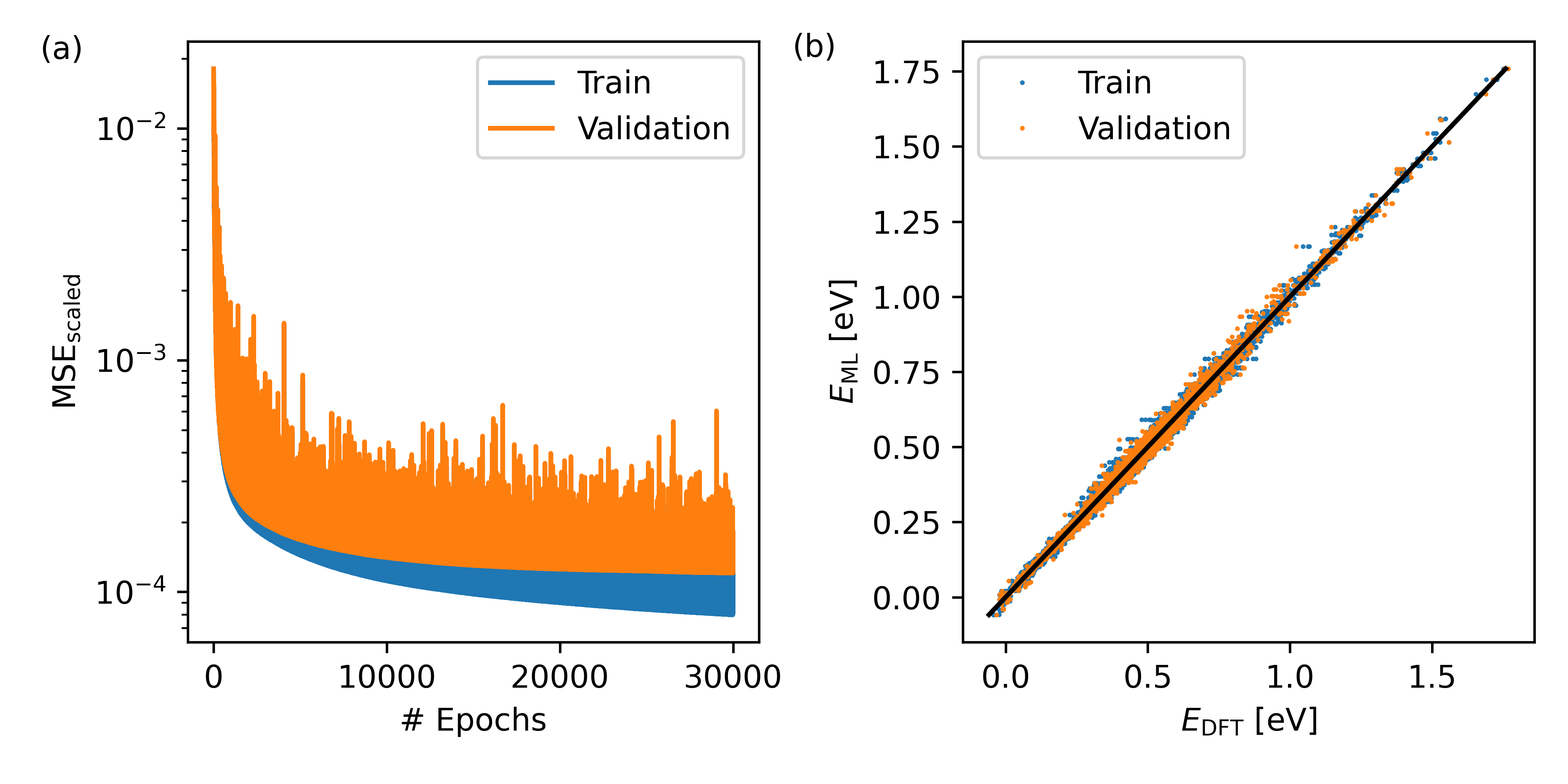}
        \caption{Training procedure of the combined 6$\times$4 and 12$\times$2 model characterized with a (a) learning curve and a (b) scatter plot comparing DFT- and ML-based energies.}
        \label{fig_sup:trainb}
    \end{figure}

    \begin{figure}[!h]
        \centering
        \includegraphics[width=0.6\textwidth]{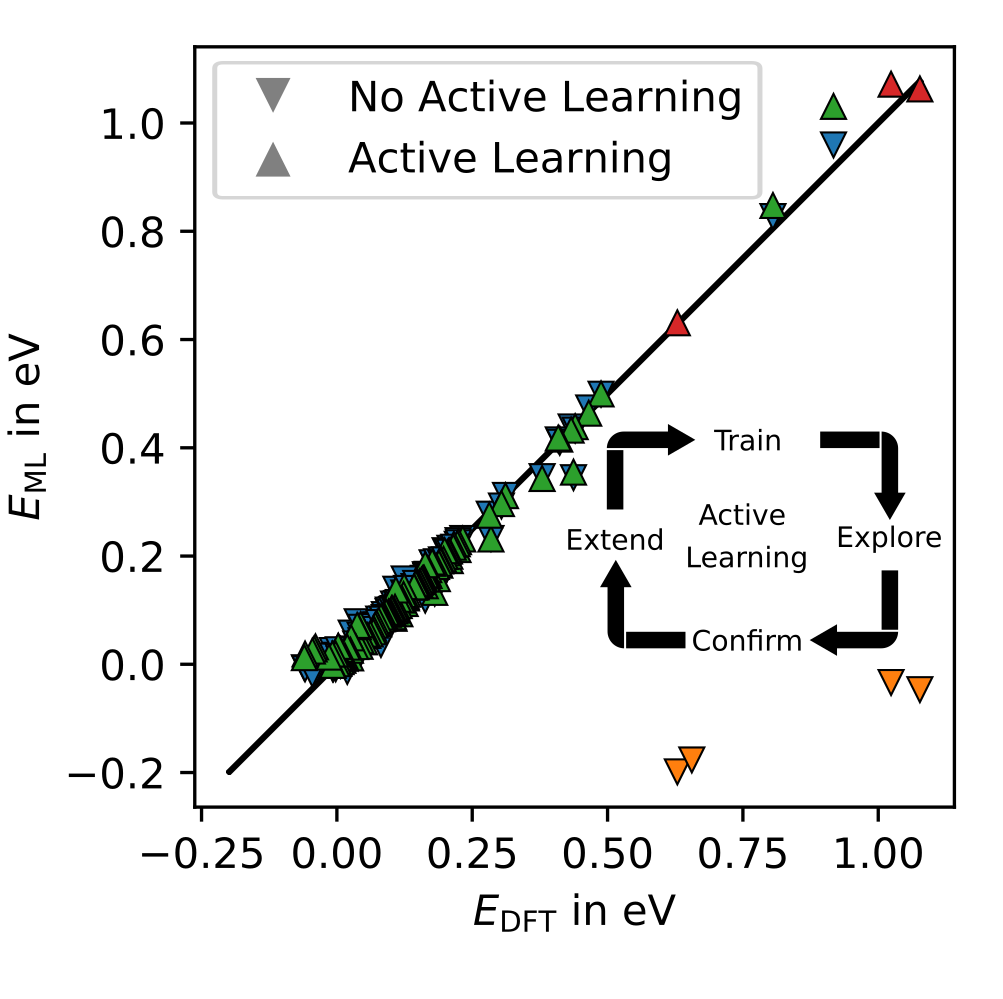}
        \caption{Comparison of test-set predictions for the model with (up-pointing triangles) and without (down-pointing triangles) iterative active learning. Blue and green data points were not added to the training data, orange and red have been used to extend training data. The inset schematically displays the active learning loop}
        \label{fig_sup:tvt}
    \end{figure}

\section{Exhaustive \VO-configurations in the 6$\times$4-cell} \label{sec_sup:conf}

    \begin{figure}
        \centering
        \includegraphics[width=\textwidth]{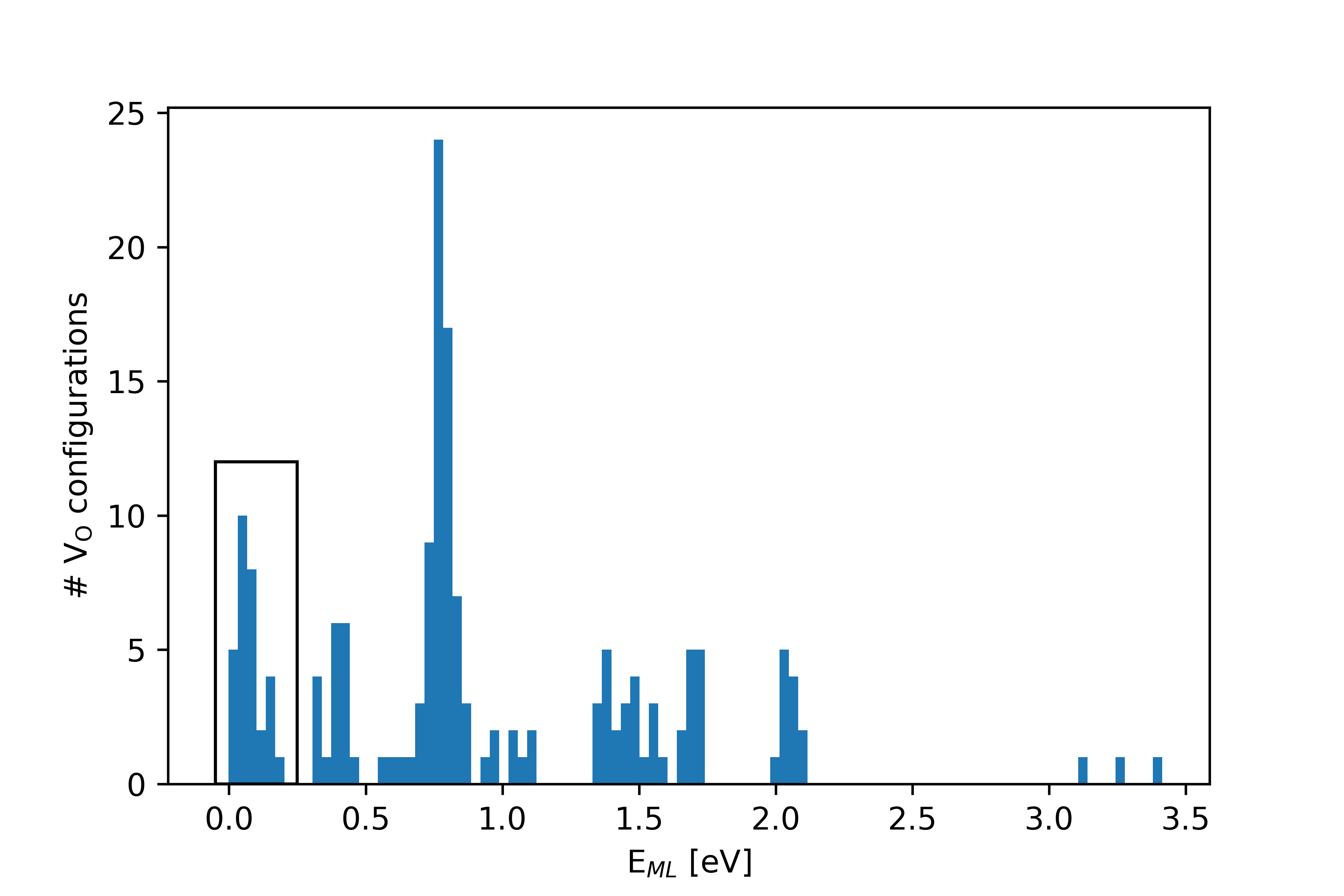}
        \caption{Distribution of ML-predicted lowest energies within all possible \VO-arrangements in the $6\times 4$-cell. Only energies of the optimized polaronic structure for each \VO-configuration are displayed. The highlighted cluster in the bottom left corresponds to the in-depth optimized configurations displayed in Figure \ref{fig_sup:low_confs}.}
        \label{fig_sup:dist}
    \end{figure}
    During the active learning scheme, we exhaustively predicted optimal polaron configurations in all possible oxygen vacancy configurations in the $6\times 4$-cell.
    Figure \ref{fig_sup:dist} shows the energy distribution of the most favorable configurations as predicted by the ML model. 
    The marked region at the lowest energies shows the most interesting configurations and is comprised of all configurations that feature one oxygen vacancy per [001] bridging oxygen row.
    Less stable configurations were not explored as thoroughly during active learning and might contain less accurate predictions, or some configurations, which were not reproducible at the DFT level.

    \begin{figure}
        \centering
        \includegraphics[width=0.95\textwidth]{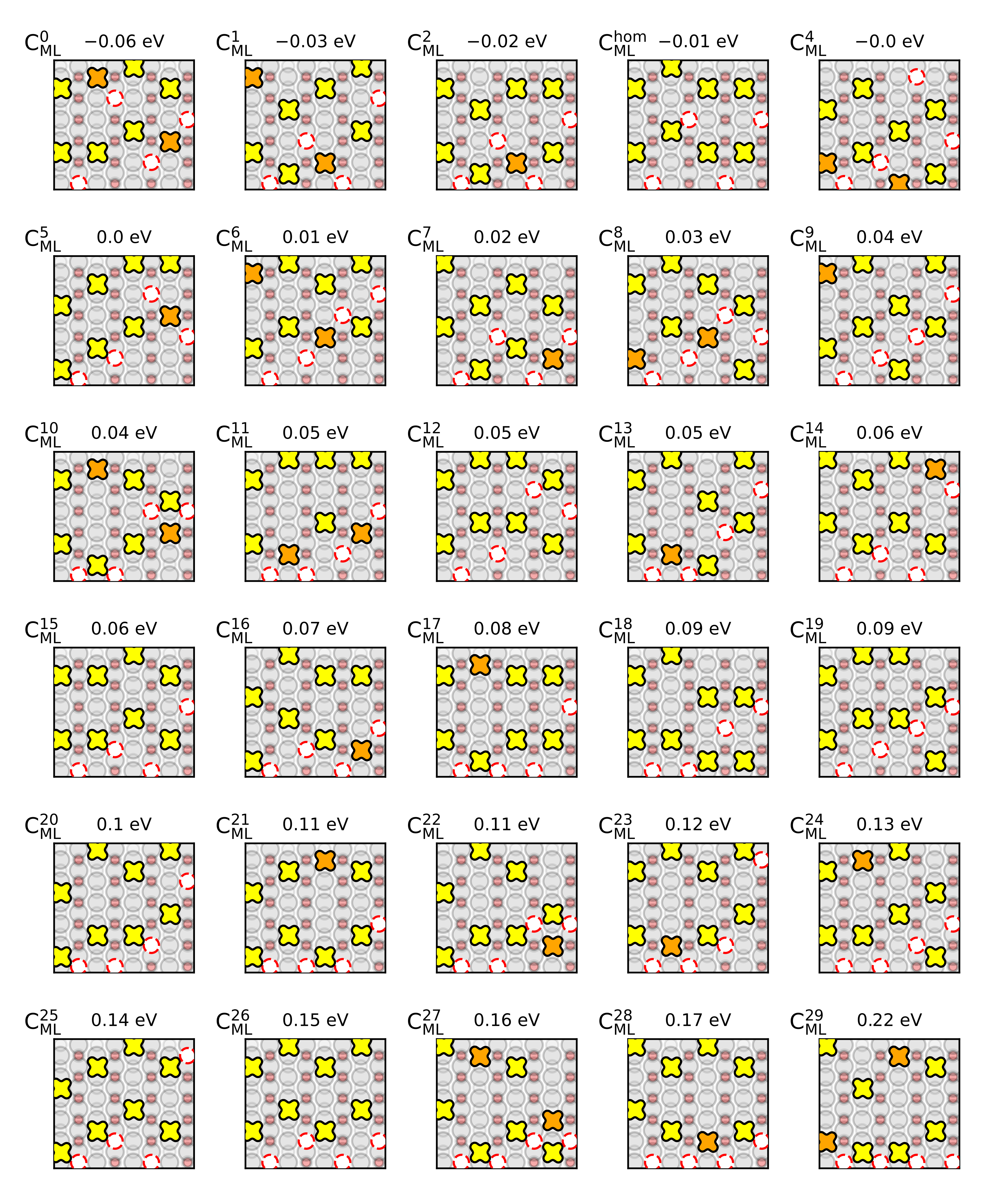}
        \caption{Schematic representation of the ML-determined optimal polaron-\VO\ configuration. 
        The 30 most stable \VO-configruations are displayed as determined by simulated annealing of the polaron configuration in a fixed \VO-configuration. 
        \VO\ are displayed as dashed red circles, surface polarons in orange, and subsurface polarons in yellow.
        The DFT energies are given relative to the random polaron homogeneous \VO\ distribution.
        The number in the label indicates the stability rank.}
        \label{fig_sup:low_confs}
    \end{figure}

    \begin{figure}
        \centering
        \includegraphics[width=0.5\textwidth]{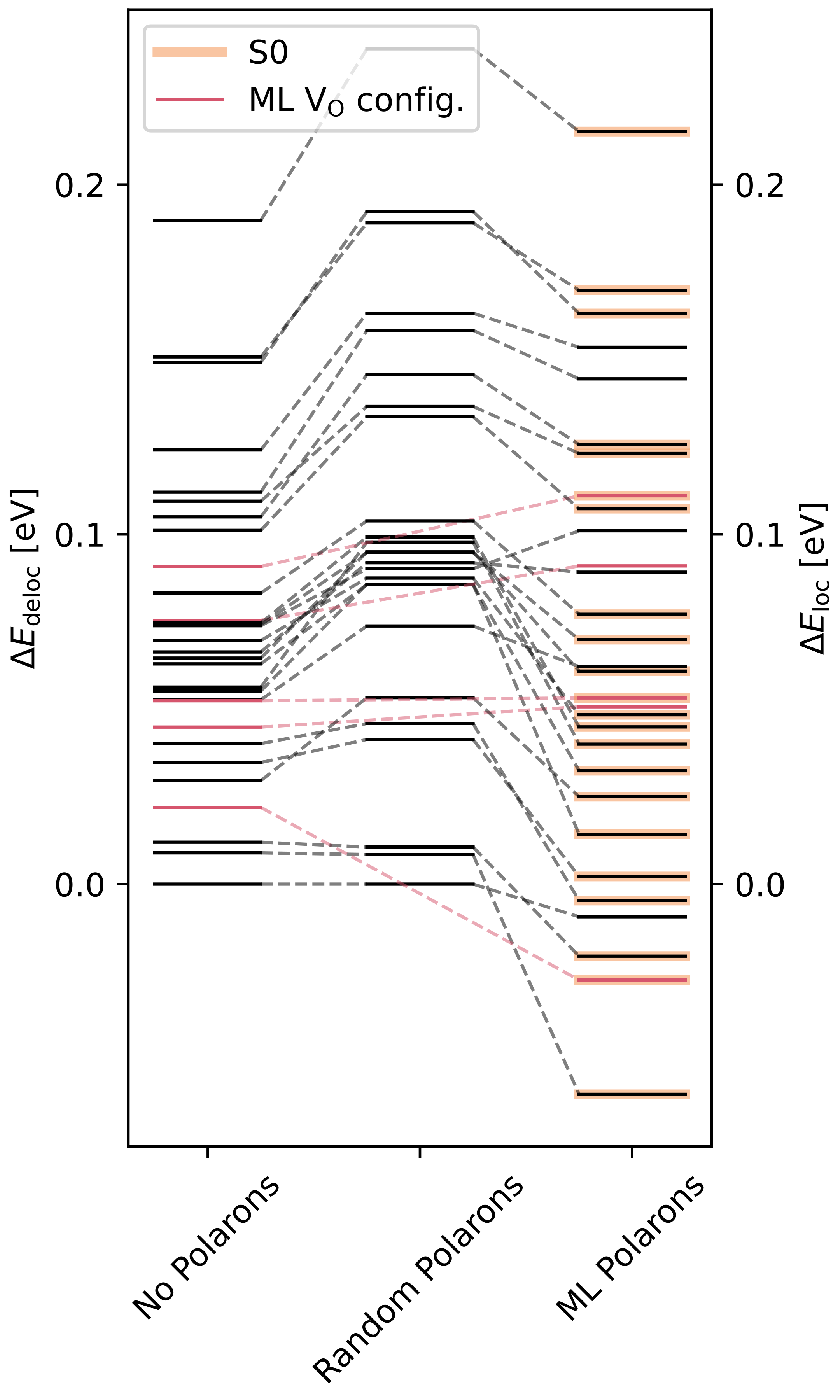}
        \caption{Changes of energies for all \VO\ configurations in the low energy cluster by different treatments of polaron-defect interactions.
        The ML Polarons column corresponds to the configurations displayed in Figure \ref{fig_sup:low_confs}, where the most stable configuration in the bottom of the ML Polarons column is $\mathrm{c}_{\mathrm{ML}}^0$ and the highest energy level belongs to $\mathrm{c}_{\mathrm{ML}}^{29}$.
        The other configurations are labeled sequentially.
        Configurations containing S0 polarons are highlighted in orange.
        The left energy scale $\Delta E_{\mathrm{deloc}}$ labels the No Polarons column.
        The other two polaronic columns are labeled by the right energy scale $\Delta E_{\mathrm{loc}}$.
        }
        \label{fig_sup:energies}
    \end{figure}

    In Figure \ref{fig_sup:low_confs}, we collect the best polaron configuration for all \VO-configurations in the low energy cluster, with their respective DFT energy. 
    Figure \ref{fig_sup:energies} collects the energy changes due to different treatments of polaron-defect interaction in the low energy \VO\ configuration cluster.
    Some notable features here are the presence of \TiSo-polarons (marked in orange) as compared to the reference dataset, where most stable configurations usually only consisted of \TiSi-polarons (marked in yellow).
    The presence of \TiSo-polarons is usually associated with two oxygen vacancies aligned along [1$\bar{1}$2] or [1$\bar{1}$0], which stabilizes the excess charge at a surface Ti$_\mathrm{5c}$ site.
    Subsurface polarons are most stable in the homogeneous ground state, in particular, when placed symmetrically and diagonally aligned between two oxygen vacancies in adjacent rows.

\newpage
\section{Experimental Autocorrelation Function} 

\begin{figure}
    \centering
    \includegraphics{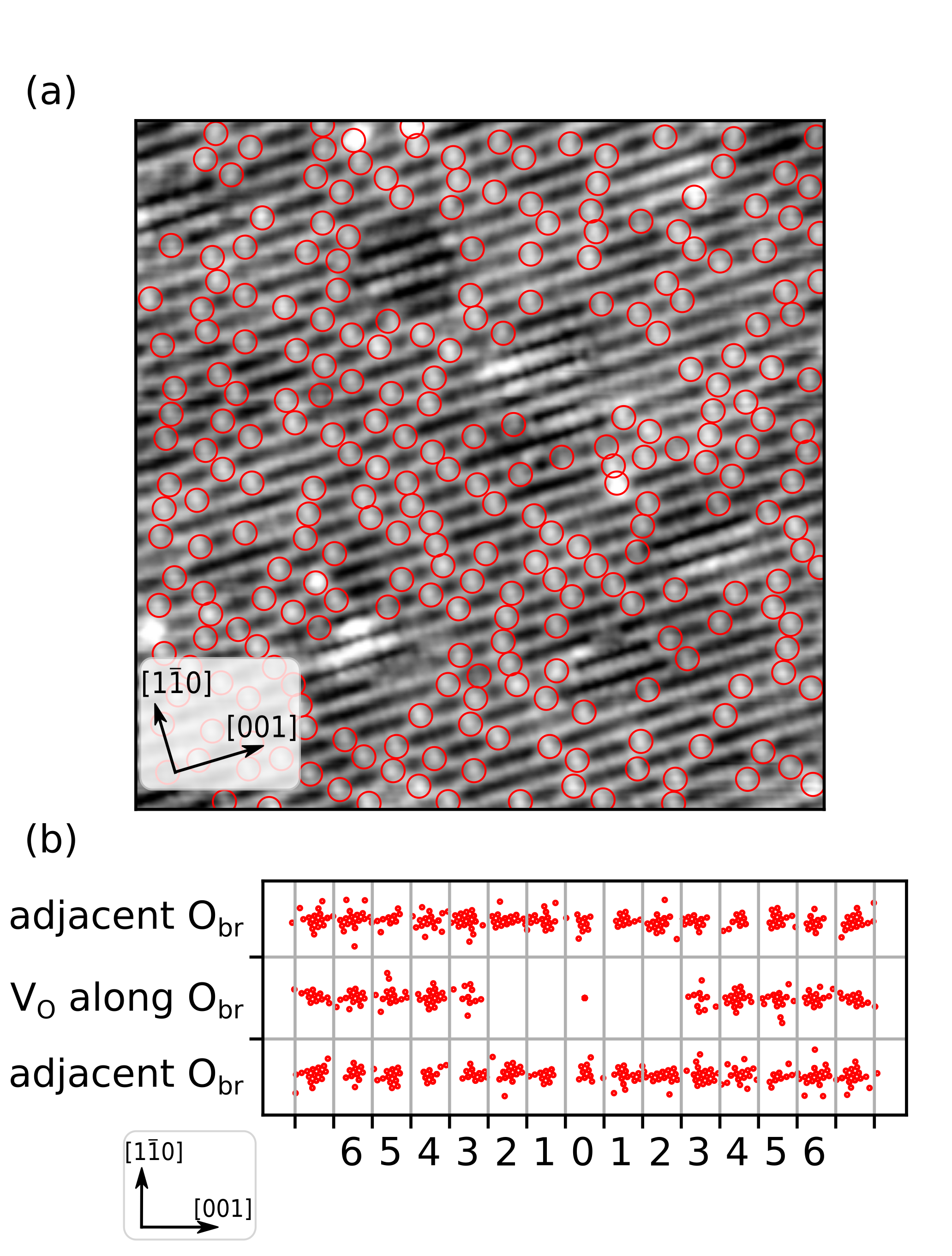}
    \caption{Experimental autocorrelation function as extracted from Fig. 1b. 
    (a) Detected \VO\ positions marked by red circles in STM measurement. 
    (b) Experimental autocorrelation function of the \VO\ positions along a single and in the adjacent [001]-aligned \obr\ rows.}
    \label{fig:autocor}
\end{figure}

    To quantify the attraction and repulsion of \VO s on the rutile TiO$_2$(110) surface, we calculated the autocorrelation function of \VO\ positions as extracted by STM measurements. 
    First, we identified the \VO\ positions in the STM image (Figure \ref{fig:autocor}a). 
    Then, we calculated the average number of \VO s around each detected \VO\ position. 
    This was done by overlaying copies of the \VO\ positions on top of each other, such that each \VO\ was placed at the origin once (Figure \ref{fig:autocor}b). 
    The resulting grid-like distribution was subdivided into individual grid cells, and the number of \VO s in each cell was counted.